%% file: main.tex
\documentclass[sigconf]{acmart}

\AtBeginDocument{%
  }

\setcopyright{none}
\renewcommand\footnotetextcopyrightpermission[1]{}
\graphicspath{{figs/}{figures/}{pictures/}{images/}{./}} 

\usepackage{tabu}                      
\usepackage{booktabs}                  
\usepackage{lipsum}                    
\usepackage{mwe}                       
\usepackage{colortbl}
\definecolor{lightgray}{RGB}{212,212,212}
\usepackage{mathptmx}                  
\usepackage{stfloats} 
\usepackage{caption}
\usepackage{subcaption}
\usepackage{relsize} 
 
\usepackage{booktabs} 
\usepackage{color} 
\usepackage{xcolor}
\usepackage{array} 

\usepackage{fancyhdr}

\usepackage{url}
\usepackage{hyperref}

\newcommand{\eg}{{e.g.,}\xspace}
\newcommand{\ie}{{i.e.,}\xspace}
\newcommand{\etal}{{et al.}\xspace}
\newcommand{\bpstart}[1]{\noindent{\textbf{#1.}}}
\newcommand{\technique}[1]{\paragraph{\textbf{#1}}}

\newcommand{\component}[1]{%
  \raisebox{-0.16cm}{\includegraphics[height=0.05\linewidth]{figures/#1}}%
}

\title[A Unified Analysis of Interaction Authoring Tasks in Data Visualization]{Intents, Techniques, and Components: a Unified Analysis of Interaction Authoring Tasks in Data Visualization}


\setcopyright{none} 
\renewcommand\footnotetextcopyrightpermission[1]{} 
\makeatletter
\def\@copyrightspace{\relax}
\makeatother

\title{Intents, Techniques, and Components: a Unified Analysis of Interaction Authoring Tasks in Data Visualization}

\author{Hyemi Song}
\affiliation{%
  \institution{University of Maryland College Park}
  \city{CollegePark}
  \country{USA}
}
\author{Sai Gopinath}
\affiliation{%
  \institution{University of Maryland College Park}
  \city{CollegePark}
  \country{USA}
}
\author{Zhicheng Liu}
\affiliation{%
  \institution{University of Maryland College Park}
  \city{CollegePark}
  \country{USA}
}

\begin{document}

\thispagestyle{empty} 
\pagestyle{empty} 

\begin{teaserfigure}
  \includegraphics[width=\textwidth]{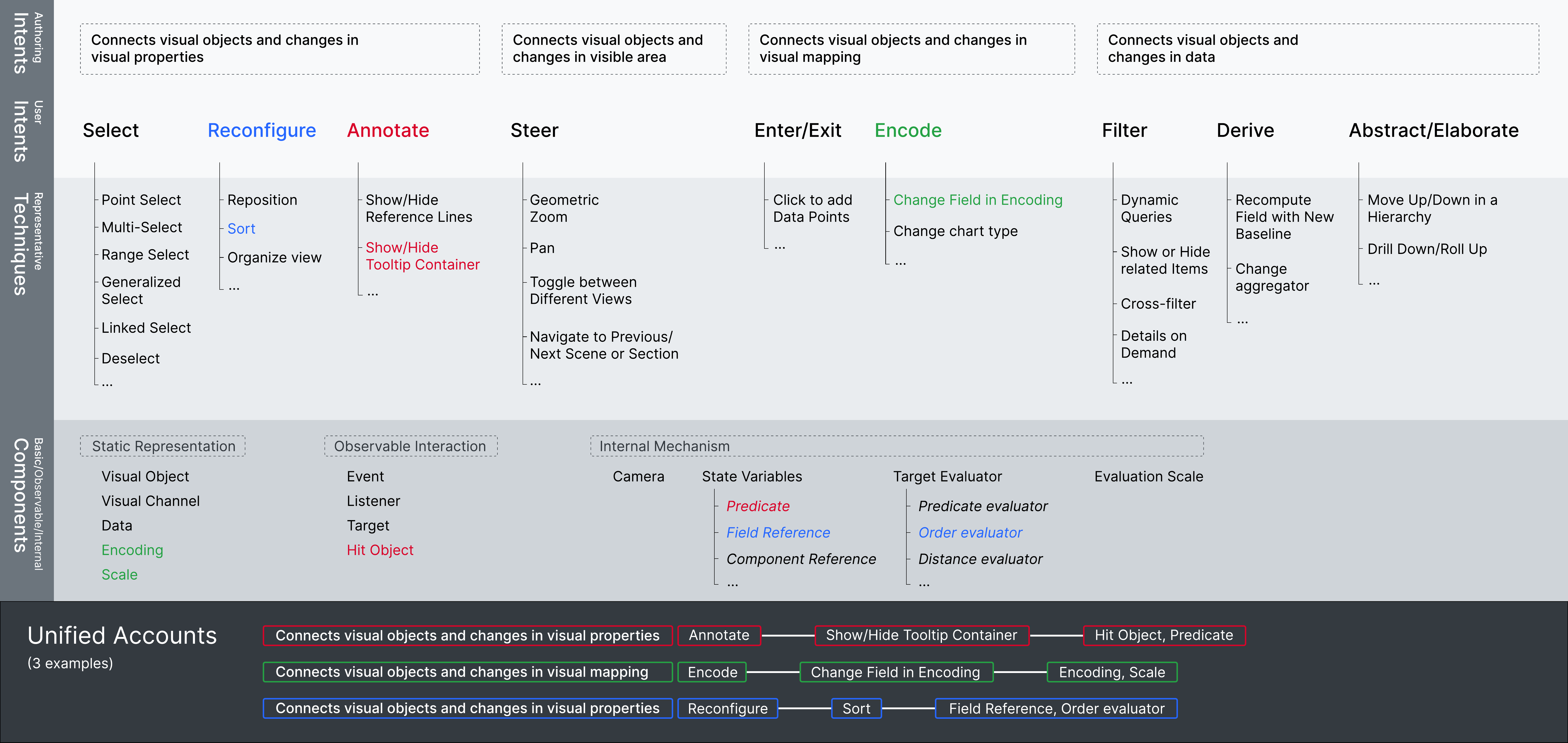}
  \caption{A summary of the categories in our unified analysis of the interaction design space across three levels of abstraction: intents,
techniques, and components. Three examples are highlighted using different colors to show how a technique embodies a high-level
intents, and what low-level components are involved in the implementation of the technique.}
  \label{fig:teaser}
\end{teaserfigure}

\begin{abstract}
  There is a growing interest in designing tools to support interactivity specification and authoring in data visualization. To develop expressive and flexible tools, we need theories and models that describe the task space of interaction authoring. Although multiple taxonomies and frameworks exist for interactive visualization, they primarily focus on how visualizations are used, not how interactivity is composed. To fill this gap, we conduct an analysis of 592 interaction units from 47 real-world visualization applications. Based on the analysis, we present a unified analysis of interaction authoring tasks across three levels of description: intents, representative techniques, and low-level implementation components. We examine our framework's descriptive, evaluative, and generative powers for critiquing existing interactivity authoring tools and informing new tool development.
\end{abstract}

\maketitle

\noindent {\small © 2025 All Authors. Version 3 (v3) supersedes v1 and v2, which used different license and acknowledgment terms. Please do not use v1 or v2.}

\input{sections/1._intro}

\input{sections/2._related_work}
\input{sections/3._method}
\input{sections/4._intent}
\input{sections/5._technique}
\input{sections/6._component}
\input{sections/7._design_patterns}
\input{sections/8._discussion_and_conclusion}
\input{sections/9._conclusion}

\begin{acks}
This research was conducted as part of the requirements for Hyemi Song’s Master’s degree in Computer Science at the University of Maryland, College Park, from Spring 2022 to Summer 2023. This paper was peer-reviewed and invited for fast-track submission to the Computer Graphics Forum (CGF) journal via EuroVis 2023.
\end{acks}

\bibliographystyle{ACM-Reference-Format}
\bibliography{main}

\section*{Contact}
For correspondence, please contact the first author (hsong02@cs.umd.edu) at the University of Maryland College Park.

\end{document}

%% file: sections/1._intro.tex

\section{Introduction}
As the support for creating static visual representations of data has matured in visualization grammars and authoring tools \cite{liu_data_2018,ren_charticulator_2018,satyanarayan_lyra_2014,tsandilas_structgraphics_2020,wang_falx_2021}, there is a growing interest in enabling easy \textit{interaction authoring}--- adding interactive behavior to static visualizations---
through declarative languages and interactive systems 
\cite{satyanarayan_vega-lite_2016,zong_lyra_2020,latif_kori_2022,sultanum_leveraging_2021,morth_scrollyvis_2022}. To  understand the expressiveness of extant interaction authoring systems and to inform the design of new languages and tools, it is important to have theories and models that describe the task space of interaction authoring in data visualization. While previous research has provided taxonomies and frameworks \cite{yi_toward_2007,heer_interactive_2012,munzner_multi-level_2013,gotz_characterizing_2009} to describe tasks in interactive visualization, they primarily focus on how visualizations are used for analytical or communication purposes, rather than how interactivity is composed. In this paper, we address this gap by providing a unified framework that describes interaction authoring tasks at multiple levels of abstraction.

To understand what level of abstraction means in the context of interaction authoring,
consider the scenario
where an author wants to add interactivity to a static visualization consisting of multiple line graphs showing stock prices of various companies over a period of time. At the lowest level, the author may specify that when the user draws a path, the visualization should update to show only the line marks with trends similar to the path drawn. This concrete specification may be composed using a programming language. At a higher level, this specification can be more abstractly described as ``query sketch'' \cite{wattenberg_sketching_2001}, which may be provided as a standard template in an authoring tool with graphical user interfaces. At the highest level, the author may have the option to just state that she wants to add filtering support to a recommendation system, which can generate applicable techniques and corresponding low-level implementations for her to choose from. 

To analyze such multifaceted interaction authoring tasks, we first identify three levels of abstraction in interaction authoring: intents, techniques, and components.
We then classify 592 interaction units from 47 real-world visualization applications into categories at the intent and component levels. With the resulting intent and component taxonomies, we identify representative techniques for each intent category and required components for each technique in a unified analysis.

Through the analysis, we make the following contributions:
\begin{itemize}
    \item We reinterpret and consolidate existing taxonomies on user intent. Based on the refined user intents, we define four types of \textit{authoring intent}: Connects visual objects and changes in visual properties, Connects visual objects and changes in visible area, Connects visual objects and changes in visual mapping, Connects visual objects and changes in data.
    \item We consolidate techniques mentioned in existing taxonomies and extend the lists of techniques based on example analysis.
    \item We identify a set of low-level primitives for composing interactivity, organized into three categories: representation components (\eg visual object, channel, encoding, scale), observable interaction components (\eg event, hit object, target), and internal components (\eg camera, state variable, target evaluator).
    \item We contribute a unified framework of the task space of interaction authoring across three levels of abstraction: intents, techniques, and components (Figure \ref{fig:teaser}). For each type of user intent under an authoring intent, we identify representative interaction techniques; for each technique, we describe how the desired behaviors can be achieved through the low-level components. 
\end{itemize}
We discuss the descriptive, evaluative, and generative powers of our analysis. In particular, we discuss the implications of our work on the design and evaluation of interaction authoring tools.
As part of the third contribution, we build an interactive web application that enables visualization researchers to collect real-world examples collaboratively. The website can be found in our supplementary material~\footnote{\href{https://interactionauthoring.github.io/app/}{supplementary materials}}.

%% file: sections/2._related_work.tex
\section{Background}
\label{sec:background}
\subsection{Interaction Authoring}
\textit{Interaction} is a frequently used term in a wide range of literature on HCI and Visualization. These two survey papers \cite{hornbaek_what_2017, dimara_what_2019} introduce seven concepts (\ie Dialogue, Tool Use, Optimal Behavior, Transmission, Embodiment, Experience, and Control) of interaction. Among the concepts, \textit{Dialogue} \cite{hornbaek_what_2017, dimara_what_2019} defines interaction as a cycle of communication acts channeled between human (via perception and action) and a system (via input and output). By drawing from Norman's seven stage model \cite{sedig_interaction_2013}, this concept is emphasized with \textit{mapping} that a user seeks for a way to achieve their intention with tasks that a user translates their goals to a intended system state. In addition, tools or models for constructing interactions \cite{leiva_enact_2019, blouin_interacto_2021, martin_causette_2022, beaudouin-lafon_instrumental_2000} introduce how they define interaction and how their interaction building system works. Leiva \etal \cite{leiva_enact_2019} presents a prototyping tool for touch-based mobile interactions. They define interaction as a set of rules to associate user inputs with system outputs. To elaborate, user inputs (\ie touches, movements, expressions, voice) captured by devices are translated into system outputs (\ie UI feedback, underlying operations), resulting in visual outputs through feedback from corresponding operations performed by the underlying system such as databases or servers. To construct such interactions, the tool user employs a first-class object called touch events. the user can manipulate touchable UI elements or program functions to establish interactive behaviors (i.e., touch-stamp, pinch, drag, touch), thereby establishing connections between input and output properties (\ie input: touch, output: the position change of an item). Similarly, other systems define interactions in the same manner: to support dialogues between a system and its users who engage with a UI to control the underlying system \cite{blouin_interacto_2021}, to establish causal relationships, such as input events instigating state changes or the propagation of values \cite{martin_causette_2022}, and to describe methods to integrate techniques by delineating the feedback from interactions based on the users' perspective \cite{beaudouin-lafon_designing_2004}. While the existing body of literature and systems depict interaction and the process of constructing it as connecting humans and systems across various stages from input to output, this concept is not limited to data visualization.

The survey paper \cite{dimara_what_2019} argues the importance of involving \textit{Data entity} and \textit{Data-related intent} to the definition of interaction in visualization. For example, the entity for interactive data visualization would be data, computer, and human. It is because visualization is for knowledge construction or sense-making data rather than the easiness or intuition of using a system. In addition, visualization, as a data interface, enables human to manipulate an intangible data sources, which can be linked to data-related user intent. This data-oriented interpretation of interaction is also seen in visualization authoring tools and languages \cite{bostock_d3_2011,satyanarayan_vega-lite_2016,lu_automatic_2021,zong_lyra_2020, satyanarayan_reactive_2015}.

Upon reviewing the literature, we discovered that connecting tasks at multi-level abstraction are crucial for interpreting interaction authoring. Consequently, this paper considers \textit{Interaction Authoring in Visualization} as a process of connecting a user with a system by integrating techniques, system components, and data throughout the entire interactive visualization pipeline.

\vspace{-2mm}
\subsection{Existing Frameworks and Taxonomies in Interactive Visualization}
Describing human behavior at multiple levels of abstraction is a cornerstone of activity theory \cite{brown_perspectives_1999,engestrom_activity_1999}, which has been informing research and design in human-computer interaction \cite{kuutti_activity_1996,nardi_activity_1996,nardi_context_1998}. Gotz and Zhou \cite{gotz_characterizing_2009} adopt activity theory to describe four levels of abstraction in visual analytics. Pike \etal \cite{pike_science_2009} identify high-level and low-level elements of interaction in terms of both user tasks and visualization features. Roth \cite{roth_empirically-derived_2013} develop a multi-tier taxonomy for cartographic interaction primitives at multiple levels: goals, objectives, operators, and operands. Brehmer and Munzner \cite{munzner_multi-level_2013} present a multi-level typology that describes the why, how, and what dimensions of user task in visualization. These works emphasize the importance of having a holistic understanding across multiple levels of abstraction, but they all focus on tasks in \textit{using} a visualization, 
and does not address the \textit{creation} of interactive visualizations. 

Taxonomies and models of interaction in visualization also exist at individual levels. Yi \etal \cite{yi_toward_2007} introduce a taxonomy of user intent, by raising a need to connect user goals with interaction techniques. Pike \etal \cite{pike_science_2009} also discuss interactions at the high-level (user intents) and the low-level (techniques). In addition, Heer and Shneiderman \cite{heer_interactive_2012} present categories of interactive dynamics capturing important user tasks. The works offer an understanding of the design space of interactive visualization at the user intent or techniques level.

The early literature also introduced taxonomies at the individual level. Dix \etal \cite{dix_starting_1998} present four classes of tasks for incorporating simple interaction into static visualization, Buja \etal \cite{buja_interactive_1996} define three classes of view manipulation along with corresponding tasks for high dimensional data visualization, and Shneiderman \cite {shneiderman_eyes_2003} suggests seven tasks based on data types. However, these taxonomies only cover different classifications at the technique level, necessitating the consolidation of existing taxonomies. \cite{schulz_design_2013, rubab_examining_2021} At the implementation level, Chi \etal \cite{chi_operator_1998} offer a taxonomy based on operators and operands. Vega-Lite \cite{satyanarayan_vega-lite_2016} presents a declarative grammar that describes how to compose interactivity with primitives; however, they focus more on the grammars' capabilities rather than linking different levels. Animated Vegalite \cite{zong_animated_2022} discusses unified interaction and animation tasks based on their similarities at high and low levels of abstraction. Reactive Vegalite \cite{satyanarayan_reactive_2015} presents a comprehensive data model demonstrating the entire data transformation process, from user input to output, in interactive data visualization. However, these two works primarily focus on lower levels, minimizing the involvement of human activities such as user intent.

Our work seeks to reinterpret and link these separate single-level accounts in a unified analysis focusing in interaction authoring tasks, rather than communication. In trying to connect these multiple levels of abstraction, we also modify some of these classifications to derive new concepts and taxonomies for interaction authoring.

%% file: sections/3._method.tex
\vspace{-2mm}

\section{Choice of Abstraction Levels}
\label{sec:approach}
We decide to perform analysis on interaction authoring tasks at multiple levels of abstraction because current literature has pointed out that a single level of abstraction is insufficient to capture the rich semantics of visualization activities in terms of motivation, context, goals and operations \cite{pike_science_2009,roth_empirically-derived_2013,munzner_multi-level_2013,gotz_characterizing_2009}. The scenario in the introduction illustrates the importance of supporting interaction authoring at multiple levels of abstraction. Although a few multi-level visualization frameworks exist (\eg Brehmer and Munzner \cite{munzner_multi-level_2013}, Gotz and Zhou \cite{gotz_characterizing_2009}), we did not closely follow their level characterizations because of our distinct focus on interaction authoring.  In this section, we define the three levels of abstraction (\ie intent, technique, and component) in our analysis and explain why they are chosen by situating them in the context of existing literature. 

\bpstart{Intents}
 In our framework, ``intents`` include ``authoring intent`` and ``user intent``. Firstly, ``authoring intent'' is what an author wants to achieve when authoring interaction. We define ``authoring intent`` based on ``user intent`` from existing literature, and the definition of interaction authoring in the section \ref{sec:background}. Yi \etal \cite{yi_toward_2007} emphasize the importance of linking ``user intent`` with techniques, demonstrating the impact of ``user intent`` throughout the data exploration process. As a result, we regard their taxonomy as a primary reference when consolidating ``user intent``. They elaborate that different interaction designs can be interpreted to achieve the same goal, called ``user intent``. For example, consider the following two designs: (1) clicking on a circle in a scatter plot changes its color, and (2) swiping horizontally over a bar chart on a phone fades the bars outside the swiped region into the background. Both interaction designs serve the same intent: to select items of interest. Besides Yi \etal's taxonomy, Heer and Shneiderman's interactive dynamics categories \cite{heer_interactive_2012} and Brehmer and Munzner's \textit{how} level \cite{munzner_multi-level_2013} are also operating at this level of abstraction for ``user intent``. The former presents a more extensive list of categories, while the latter not only addresses the user intent but also examines the relationships between user intents and other levels. Taking these interpretations into account, we define ``authoring intent'' that is to seek a task to translate these ``user intents`` to desired system elements. For example, to enable users to select a bar, an author wants to connect a bar and corresponding changes in its visual properties by programming.

\noindent\bpstart{Techniques} In our framework, techniques are to achieve a user intent, which is ultermately for the goal of fulfulling authoring intent. Techniques often tightly coupled with the type of items being manipulated. For instance, the two designs in the previous paragraph are examples of two different techniques: point select and range select. The former is about selecting a single item, while the latter is about selecting items within a specific pixel range. For the same reasons associated with the intents, we regard the taxonomies by Yi \etal \cite{yi_toward_2007} and Heer and Shneiderman \cite{heer_interactive_2012} as our primary references. These taxonomies feature representative techniques for each high-level user intent. We consolidate their techniques and add more techniques discovered while analyzing examples.



\noindent\bpstart{Components} Finally, components are the building blocks to implement interaction designs.
We consider this level similar to the \textit{what} level in Brehmer and Munzner's typology \cite{munzner_multi-level_2013}. Their account at the \textit{what} level, however, is not detailed and does not address interaction authoring. In general, there is little work that systematically identifies the primitives in authoring interactive visualizations. A represented relevant work is Vega-Lite \cite{satyanarayan_vega-lite_2016}, which identifies \textit{what} comprises interactivity from the perspective of declarative specification. Our analysis at this level seeks to provide more details on \textit{what} inputs and outputs are involved to translate UI events into visual changes.

Given these three levels of abstraction, we provide an overview of the analysis methods and process used to identify the specific categories at each level in Section \ref{sec:method}. Section \ref{sec:intents}-\ref{sec:unified} then present the results of the identified categories at each level and a unified account of how these categories relate to each other across levels.

\vspace{-2mm}
\section{Analysis Method and Process}
\label{sec:method}
The primary goal of our analysis is to identify common categories at each of the three levels and organize these categories across levels. Similar to previous work that focuses on identifying high-level intents \cite{yi_toward_2007} and low-level primitives \cite{liu_data_2018,thompson_understanding_2020} in data visualization, 
we ground this work in the analysis of real-world interactive visualization examples.

\bpstart{Example Collection}
We collected examples based on the following considerations. First, the static visual representations should encompass a variety of chart types and designs. Second, the interactive behavior should go beyond minimal features such as showing tooltip on hover. Third, we want to sample visualizations designed for both exploratory analysis as well as storytelling. 


Based on these criteria, we identified 47 real-world interactive visualization examples from 16 online sources, including web applications (e.g., GOV$|$DNA \cite{sudox_govdna_nodate}, ESA Star Mapper \cite{interactive_esa_nodate}), news outlets (\eg the New York Times \cite{the_new_york_times_new_nodate}, the Economist \cite{the_economist_big_nodate}), digital publishers (\eg the Pudding \cite{the_pudding_pudding_nodate}, Quartz \cite{quartz_quartz_nodate}), design galleries of visualization toolkits (\eg bl.ocks \cite{popular_blocks_popular_nodate}) and systems (\eg Tableau Public \cite{tableau_tableau_nodate}, datasketch \cite{datasketch_datasketch_nodate}), commercial demos (\eg heavy.ai \cite{heavyai_1990_nodate}, Gramener \cite{gramener_gramener_nodate}), and portfolio sites of design agencies (\eg Column Five Media \cite{column_five_media_column_nodate}). 

In terms of chart design, 15 out of 47 examples consist of multiple views, which are arranged either as panels laid out in a dashboard, or as frames in a scrollytelling format, or as tabs activated using toggle buttons. In total, the example collection represents 20 different chart types: scatter plot, symbol map, heat map, choropleth map, streamgraph, bubble chart, sunburst chart, index chart, simple/grouped/stacked bar chart, (stacked) area chart, dumbbell chart, radar chart, 3D particle map, treemap, 3D area chart, network map, and line chart. 


\bpstart{Assumptions} 
Given our focus on interactivity, we assume that the static representations have been created.
If a visualization component does not change in response to user action, it is not included in the analysis. 

\begin{figure}[tbp]
 \centering
 \includegraphics[width=\linewidth]{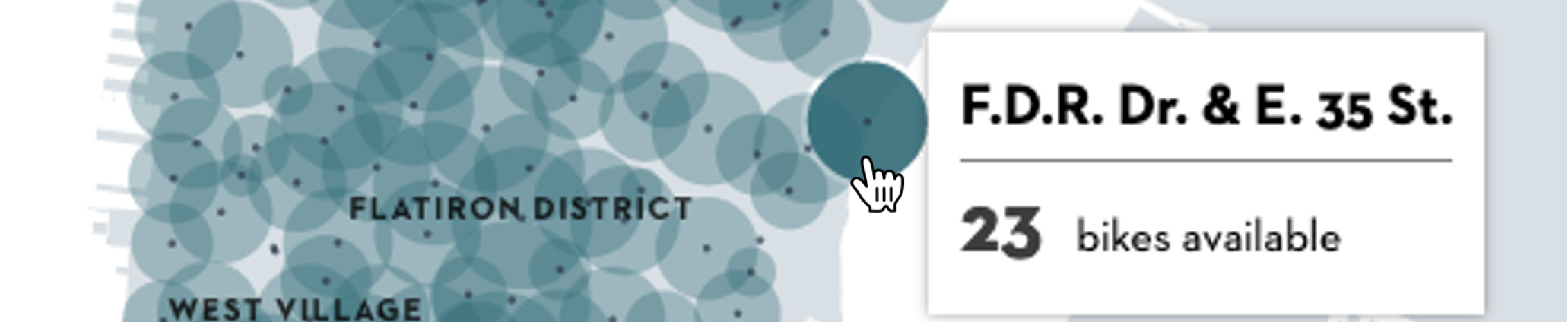}
 \vspace{-5mm}
 \caption{Hovering to highlight a circle mark and show a tooltip}
\label{fig:tooltip}
\end{figure}

\bpstart{Unit of Analysis: Event \& Target} Since many examples are quite complex, involving multiple views responding to multiple user actions, 
we chose an \textit{event-target pair} as the unit of analysis: an \textit{event} refers to a keyboard or a mouse event, and a \textit{target} refers to a set of visual objects exhibiting visible changes as a result of the event. In Figure \ref{fig:tooltip}, we have two targets for the hover event: (1) all the circles, and (2) the tooltip. This example thus is decomposed into 2 units of analysis. 
In total, we analyzed 592 units of analysis in the 47 examples.

\bpstart{Process Overview} 
We conducted the analysis in three phases. We focus on intent categories and component categories
in the first two phases, and 
identify representative techniques
in the third phase. 

\noindent\textit{Phase 1}: We first adopted a top-down approach to apply existing taxonomies to the examples.
At the intent level, we tried to assign each unit of analysis a category by consolidating Yi \etal's taxonomy \cite{yi_toward_2007}, Heer and Shneiderman's taxonomy \cite{heer_interactive_2012}, and Brehmer and Munzner's \textit{how} classification \cite{munzner_multi-level_2013}. 
At the component level, we analyzed each unit of analysis by labeling the selection primitives in Vega-Lite's interaction grammar (\eg type, predicate, domain|range, init, transforms, resolve). 
At the early stage, All three authors independently coded the examples and met to resolve disagreements. Initially, discussions on a single complex example could take an hour due to differing interpretations of intent categories and selection primitives. However, as we aligned on the concepts, we could review three examples within an hour. After the initial ten examples, two authors continued coding and involved the third only when needed for disagreements.


\noindent\textit{Phase 2}: As we made progress in the first phase, we also discovered potential issues that prevent us from directly using existing classifications for interaction authoring. Section \ref{sec:intents}-\ref{sec:components} elaborate on these issues. We thus adopted a bottom-up approach to revise and more precisely define the categories. At the intent level, 
we paid close attention to ambiguous cases, and analyzed the similarities and essential differences between related examples to revise the intent categories. 
At the component level, we identified both components that are observable to users as well as those that are internal in a program. Section \ref{sec:intents}-\ref{sec:components} provide a detailed account of the results from this phase, along with the rationales for choosing the intent and component categories.

\noindent\textit{Phase 3}: With the revised intent categories and component types, we performed labeling adjustment. After adjusting labels at the intent level, we grouped similar units of analysis and identified common techniques associated with each intent. We also identified common components used in each technique.
Section \ref{sec:unified} provides a unified account of the results from this phase. 

%% file: sections/4._intent.tex
\section{Intent Categories}
In this section, we discuss the issues and findings related to analysis at the intent level. 
\label{sec:intents}

\subsection{Issues with Directly Using Existing Taxonomies for Interaction Authoring}
As mentioned in Section \ref{sec:approach}, we used three existing taxonomies \cite{yi_toward_2007,heer_interactive_2012,munzner_multi-level_2013} to label intent. While the taxonomies were highly useful in providing conceptual scaffolding in our analysis, we did encounter issues that prevent us from directly applying these existing taxonomies to interaction authoring: focus on user intent instead of authoring intent.


\subsubsection{User Intent vs. Author Intent}
The existing taxonomies describe intent in terms of what users want to achieve. This is evident in \cite{yi_toward_2007}, where the meaning of each intent is explained in the form of ``show me ...'', and in 
\cite{heer_interactive_2012}, where the ``categories incorporate
the critical tasks that enable iterative
visual analysis''. However, to enable users to achieve their goals, authors need to first articulate what they want to achieve in terms of \textit{how the target visual objects should change}. In other words, \textit{what to connect to change the target visual objects.}
For example, in Figure \ref{fig:tooltip}
the user intent here is \textit{elaborate} using Yi \etal's taxonomy \cite{yi_toward_2007}, which describes what users want to achieve: to get more information about the station. To support this user intent, authors need to specify how the visibility and position of the tooltip should be updated based on mouse events, as well as how the tooltip content should be updated by retrieve the corresponding data. To do so, they should intend to link the input event, the tooltip, and its visual properties. The authoring intents here are very different from the user intent \textit{elaborate}.

\begin{figure}
     \centering
     \begin{subfigure}[b]{0.45\textwidth}
         \centering
         \includegraphics[width=\textwidth]{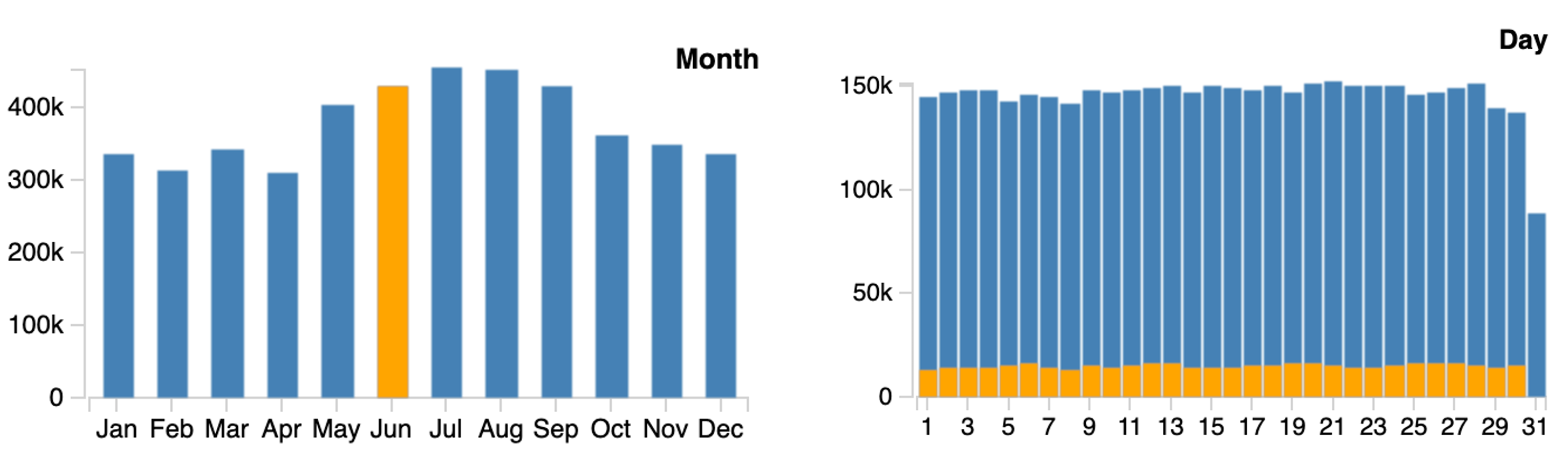}
         \caption{Distribution of visualization designs created by participants by design use frequency.}
         \label{fig:cross-filter}
     \end{subfigure}
     \hfill
     \begin{subfigure}[b]{0.39\textwidth}
         \centering
        \includegraphics[width=\textwidth]{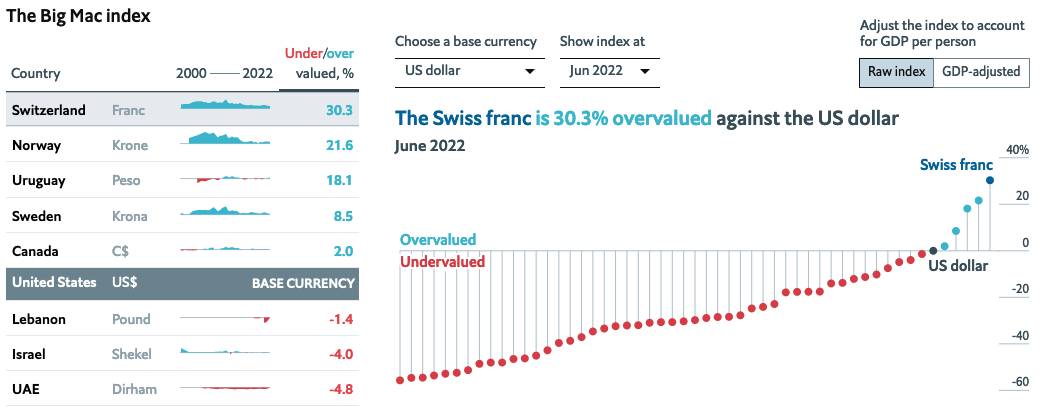}
         \caption{Distribution of participants by tool type and frequency of use.}
         \label{fig:linked-select}
     \end{subfigure}
        \vspace{-3mm}
        \caption{Interactions involving coordinated views.}
        \label{fig:coordinate}
     \vspace{-3mm}
\end{figure}

\subsubsection{Inconsistencies within the Same Taxonomy}
We also found that the categories from the same taxonomy can focus on different aspects of interactivity. Consider 
the two examples in Figure \ref{fig:coordinate}.  
Figure \ref{fig:cross-filter} \cite{liu_immens_2013} shows two bar charts, brushing over a month bar in the left chart will show the daily distribution of that month in orange in the right chart. Figure \ref{fig:linked-select} \cite{the_economist_big_nodate} shows a list of currencies in the left table, with a base currency (US dollar) colored in dark gray. The lollipop chart on the right shows all the currencies' Big Mac index against the base currency. Selecting a currency (Swiss franc) in the table will highlight the corresponding lollipop. Using Yi \etal's taxonomy, we can label the interactions in both examples as \textit{connect}: to ``highlight
associations and relationships between data items that are already
represented'' in different views \cite{yi_toward_2007}. However, if we focus on \textit{how} the targets are changing, the two examples can be categorized differently. In Figure \ref{fig:cross-filter}, the target (orange histograms) updates to show a subset of the data matching a specific month, hence it can be categorized as \textit{filter}: ``change the set of data
items being presented based on some specific conditions'' \cite{yi_toward_2007}; in Figure \ref{fig:linked-select}, on the other hand, the target (lollipop) is just a different visual representation of the same data item (Switzerland Franc), and it is differentiated  from the other lollipops using fill color, hence matching the definition of \textit{select}: ``by making items of interest visually distinctive, users can easily keep track of them even
in a large data set and/or with changes in representations'' \cite{yi_toward_2007}. 

These categories are from the same taxonomy, yet they do not focus on the same aspect of the interaction design: \textit{connect} emphasizes the relationship between the mark clicked on and the target, without describing the effects on the target; whereas \textit{filter} and \textit{select} focus on \textit{what kind of change} the target exhibits. The same issue applies to Heer and Shneiderman's taxonomy (\eg \textit{coordinate} vs. \textit{filter}).






\subsection{A Taxonomy of Intent}
Considering these two issues raised before, we decide to create our own list of categories for authoring intent and user intent by reinterpreting these existing taxonomies and analyzing examples. To do so, we need more precisely define authoring intent. When an author expresses her intent to author an interaction by connecting items, multiple aspects, derived from interactions between a user and a system, may need to be specified:
\begin{itemize}
    \item how the interaction is triggered (\eg click, scroll, or UI control)
    \item the type of target (\eg mark, view) that exhibits visual change
    \item the type of visual change (\eg reorder, data updated) in target
    \item whether the visual change triggered by interaction takes place within one view or across multiple views
\end{itemize}


Among these aspects, we argue the type of change in target is the most critical factor:
it is the most immediately observable consequence of an interaction. We thus refine the definition of authoring intent as: what an author wants to achieve in terms of \textit{the desired type of change on visual objects by connecting items}.



Our final taxonomy of authoring intent are new categories formulated based on the consolidation of user intents from existing taxonomies as well as new categories from our analysis. First of all, we collected a list of user intents related to how target object change (\eg reconfigure, encode, filter). Yi \etal's work do focus on such a perspective; hence, we mainly referred to the user intents from their taxonomy. We also made modifications based on the considerations discussed above. For example, we removed \textit{connect} and \textit{explore}, and added categories such as \textit{derive} and \textit{annotate} from Heer and Shneiderman's taxonomy and Brehmer and Munzner's typology. With the consolidated user intents, we grouped them by \textit{way of connect} as authoring intents.

Here we present the categories: Authoring Intents (AI) and related user intents, and their definitions.
We will elaborate on these categories with representative techniques and low-level semantic components in Section \ref{sec:unified}.

\noindent\bpstart{AI1: Connects visual objects and changes in visual properties}
link user inputs to related system components to evoke changes in the visual aspects of target objects, thereby realizing these user intents:

\begin{itemize}
    \item \textit{Select}: visually differentiate objects from their surroundings
    \item \textit{Annotate}: show auxiliary visual objects
    \item \textit{Reconfigure}: change the spatial arrangement of existing objects 
\end{itemize}

\noindent\bpstart{AI2: Connects visual objects and changes in visible area}
link user inputs to related system components to evoke changes in the range and viewpoint of visible area, thereby realizing this user intent:

\begin{itemize}
    \item \textit{Steer}: direct camera view towards certain objects
\end{itemize}

\noindent\bpstart{AI3: Connects visual objects and changes in visual mapping}
link user inputs to related system components to evoke changes in pairs of targets and corresponding data items, thereby realizing these user intents:

\begin{itemize}
    \item \textit{Encode}: change the visual representation of existing data items
    \item \textit{Enter/Exit}: add or delete visual objects representing data items or fields to the visualization
\end{itemize}

\noindent\bpstart{AI4: Connects visual objects and changes in data}
link user inputs to related system components to evoke changes in the underlying data engaged in targets' visual updates, thereby realizing these user intents:

\begin{itemize}
    \item \textit{Filter}: show a subset of the \textit{data items} based on certain criteria
    \item \textit{Abstract/Elaborate}: change the levels of abstraction, which determines the \textit{data items} to be visualized
    \item \textit{Derive}: compute and show new \textit{field values} from data items attached to existing objects
\end{itemize}

In some cases regarding user intents, multiple interpretations are possible. For example, when filtering a stacked area chart, the spatial arrangement of the remaining area marks will change. We categorize such cases as \textit{filter} instead of \textit{reconfigure}, because filtering is the driving force behind the observed change, and the layout mechanism is unchanged. 

%% file: sections/5._technique.tex
\section{Technique Categories}
We consolidate techniques from existing taxonomies and incorporate additional techniques from our example analysis. Our primary motivation is to reference taxonomies that focus on target change-centric perspective when defining intents. Thus, we also began the amalgamation process using Yi's and Heer and Shneiderman’s taxonomies \cite{yi_toward_2007, heer_interactive_2012}. Our initial step was to gather techniques associated with user intents from each paper. The Table \ref{tab:technique table} shows the collection. In consolidating these techniques, we further segment them into more granular levels. For example, while both papers define \textit{Select} as a user intent, they do not delve into the detailed techniques. To address this, we include the related techniques with the user intent, \textit{Select} (\ie Point Select, Multi-Select, Range Select, Generalized Select, Linked Select, and Deselect), all informed by our example analysis. More techniques can be added as needed; thus, we define a group of primary techniques in Section 8, presenting a unified analysis.

\begin{table*}[h!] 
\centering
{\footnotesize 
\arrayrulecolor{gray}
\begin{tabular}{ p{3cm}|p{5.5cm}|p{3cm}|p{5.5cm}  } 
 \hline
 \multicolumn{2}{c|}{Yi \etal \cite{yi_toward_2007}} & \multicolumn{2}{c}{Heer and Shneiderman \cite{heer_interactive_2012}} \\
 \hline
 User Intents & Techniques & User Intents & Techniques \\
 \hline
 Select &  & Select & \\
 Explore & Pan, Direct Walk & Navigate & Pan, Scroll, Zoom \\
 Reconfigure & Sort, Align/Change Baseline, Change X/Y Fields in Chart, Change Attributes Assigned to Axis, Relocate an Item, Rotate Portion of the Visualization & Sort & Order, Reorder Data by Variables \\
 Encode & Semantic Zoom, Change Visual Properties, Change Chart Types & Visualize & Assign Data Variables to Visual Objects, Add an Item \\
 Connect & Highlight Connected Items, Reveal/Expand Related Items, Hide Unrelated Items & Coordinate & Brushing and Linking \\
 Abstract/Elaborate & Geometric Zoom, Tooltip & Derive & Nesting Variables \\
  &  & Organize & Simplify Organization of Items, Switch Positions of Items, Resize a View \\ 
  &  & Record & Undo and Redo Actions, Add Metadata, Bookmark \\  
  &  & Annotate & Textual Annotation, Draw a Circle/Arrow \\  
  &  & Share & Export Views or Data, Publish a Visualization \\  
 \hline
\end{tabular}
} 
\caption{User Intents and Techniques in Yi's and Heer and Shneiderman's taxonomies} 
\label{tab:technique table} 
\end{table*} 

%% file: sections/6._component.tex
\section{Component Categories}
\label{sec:components}
In this section, we present the details on our analysis at the component level. The focus of our analysis at this level is: given a triggering mouse or keyboard event and a target, to attain the observed visual changes in the target, what visualization components are involved, and how are the components used or updated?
We first review recent literature that examines the design space of static data visualizations \cite{zhicheng_liu_atlas_2021,satyanarayan_critical_2019,liu_data_2018,ren_charticulator_2018,thompson_understanding_2020} and synthesize the components identified in these papers.
Given the static components, we then used Vega-Lite's interaction grammar to guide our top-down analysis of interactive behavior. In particular, we found the selection primitives to be powerful concepts to describe interaction behavior. Our analysis adapted and extends these concepts in two ways. First, we try to identify the components of interactivity with a focus on the input and output of each component. 
Second, the Vega-Lite does not provide an in-depth account of the role of UI controls, which are commonly used in real-world examples (26 out of 47 in our collection), and we introduced field and component references to address this issue. Below we introduce the component types uncovered from our analysis. We will elaborate on how the manipulation of these components relate to higher-level techniques and intents in Section \ref{sec:unified}.

\subsection{Basic Components of a Static Representation}
\label{sec:static-components}
\bpstart{Visual Object} The following list specifies different types of visual objects in a visualization, going up the scene graph hierarchy:
\begin{itemize}
    \item \textit{mark}: graphical elements such as shapes (rectangle, arc, line), symbols, images and texts; 
    \item \textit{glyph}: a group of marks, for example, a lollipop chart (Figure \ref{fig:linked-select}) uses glyphs consisting of one circle and a line;
    \item \textit{collection}: mark or glyph instances generated by data, for example, a simple bar chart contains one collection of bars, each bar represents a data item; collections can be nested, for example, a small multiple consists of a collection of mark collections;
    \item \textit{axis and legend}: reference objects that explain how to interpret a visualization;
    \item \textit{annotation}: auxiliary visual objects that provide contextual information or facilitate perceptual estimation;
    \item \textit{scene}: a self-contained view consisting of mark/glyph collections, axes, and legend. In a multiple coordinated view design, for example, each view is a scene. Similarly, in a tabbed interface, clicking on each tab shows one scene at a time; 
    \item \textit{section}: multiple scenes can be organized into a section, a typical practice in scrollytelling interfaces where a narrative covers multiple topics, each topic involving a set of scenes.  
\end{itemize}

\bpstart{Visual Channel} A visual object has a number of visual channels such as position, size, and color that can be used to represent data values or customize visual styles. 

\bpstart{Data} Visual objects often have attached data items, which are encoded by properties (\eg position, size) of the visual objects

\bpstart{Encoding} A declarative specification of the mapping between a data field and a visual channel.



\subsection{Observable Interaction Components}
\label{sec:observable-cmpnts}
In addition to components for static representations, we identify three components related to interaction that are visually observable: \textit{event}, \textit{hit object}, and \textit{target}. 

\bpstart{Event} Interactions are triggered by events through an input device. A mouse event consists of the following types: click, move, drag, and each event has associated parameters such as the x and y coordinates. 

\bpstart{Listener} Listeners monitor and respond to events. The most used listener is the visualization canvas, which can be implemented as an SVG element or an HTML canvas. UI controls such as buttons and dropdown menus are also common listeners.

\bpstart{Hit Object} Given a mouse event received by a canvas listener, most visualizations conduct a hit-test \cite{foley_computer_1996} to determine if any visual object intersects with mouse cursor. The result of the test, if not null, is a hit object. A hit object is usually a mark or a glyph.

\bpstart{Target} Targets are visual objects that undergo observable visual changes in response to a triggering event. In many examples, the hit object is a subset of the targets (\eg in Figure \ref{fig:tooltip}, given a collection of circles, hovering over one of circles makes it a hit object, and all the circles in the collection are the targets, whose color and opacity are updated to highlight the hit object). An interactive visualization can also have targets without a hit object present: for example, we can achieve the same effect in Figure \ref{fig:tooltip} by choosing a station from a drop-down menu.

\subsection{Internal Components}
Our analysis has identified the following components related to the internal working mechanisms of an interactive visualization. In contrast to the observable interaction components, these components have no visual representations in the user interface. 

\bpstart{Camera} A component providing a vantage point of the visual representation. A camera has a number of configurable attributes, such as focus point (center of camera), field of view (extent visible through the camera), zoom level, and rotation.   In web-based 2D visualizations, the camera is often implemented as the SVG view box \cite{mdn_web_docs_viewbox_nodate} that can pan and zoom; in 3D visualizations, perspective cameras are used widely \cite{sciencedirect_engineering_topics_perspective_nodate}.

\bpstart{State Variables} An interactive visualization often requires state variables to configure and manage its state. Events update the values of these variables, which in turn trigger visual changes in target objects. Below is a list of exemplar state variables. 

\begin{itemize}
    \item \textbf{Predicate} An expression consisting of a variable, an operator, and an operand value. The variable is usually a data field.
    For example, we can define a predicate \textit{country} $=$ ``\textit{USA}'' where the operand is a single data value and the operator is \textit{equal}; in another predicate,  \textit{country} $\in$ [``\textit{USA}'' ``\textit{Canada}'', ``\textit{Mexico}''], the operand is a list of values, and the operator is \textit{in}. In some cases, the variable can be a visual property, \eg \textit{x\_position} $\in$ [25, 50] if we want to do a collision check. A predicate can be evaluated to either TRUE or FALSE. They can thus be used to either retrieve data items, or test if a target satisfies a condition.
    \item \textbf{Field Reference} Sometimes a visualization needs to keep track of the data field used to manipulate target objects. For example, in a bar chart where the bars can be sorted by different data fields, the visualization needs a \textit{sortBy} variable to keep track of the data field determining the order.
    \item \textbf{Component Reference} Direct references to visualization components are useful for navigational and presentation purposes. For example, in a scrollytelling interface where multiple scenes are included in the same page and shown at a predefined cadence, an state variable \textit{currentScene} is necessary to decide which scene to display, and this variable can be bound to UI navigation controls like breadcrumbs as well as page view port. 
\end{itemize}



\bpstart{Target Evaluator} A target evaluator is a function that takes a list of visual objects as input, and outputs numeric or Boolean values for each object based on the visual object's data or visual properties. A target evaluator often involves one or more state variables. Below is a list of exemplar target evaluators:
\begin{itemize}
    \item A \textit{predicate evaluator} tests if visual objects satisfy a given predicate. For example, given a bar chart where each bar represents a country and its height represents the population of each country, we can define an evaluator  that tests each bar against the predicate \textit{country} $=$ ``\textit{USA}'', and the function will return a list of Boolean values, with TRUE for the bar representing USA only. 
    \item An \textit{order evaluator} sorts a list of visual objects based on a sorting criteria, which can be expressed as a field reference, and returns a list of object indices.  
    \item A \textit{distance evaluator} computes the pairwise similarity or distance between two objects. In a fisheye view, for example, given a hit object, we can define an evaluator that returns the index distance between each target and the hit object. 
    
\end{itemize}


\bpstart{Evaluation Scale} A special type of scale specifying how to update target visual objects based on target evaluation results. Unlike a regular scale which takes data values as input, an evaluation scale takes results from a target evaluator as input. 
In the bar chart example above involving a predicate evaluator, the bar satisfying the predicate can be assigned a fill color different from the other bars. 

%% file: sections/7._design_patterns.tex
\section{Intents, Techniques, and Components: A Unified Account}
\label{sec:unified}
In this section, we present a unified account of representative interaction techniques in our example collection, grouped by authoring intent and user intent in each subsection, with the low-level operation mechanisms described using static and interaction components. 
For each technique, we indicate whether it is primarily for a single view (S) or multiple views (M). We also use colored icons to highlight the components involved in each technique. When a component is in square brackets (\eg [\raisebox{-0.16cm}{\includegraphics[height=0.05\linewidth]{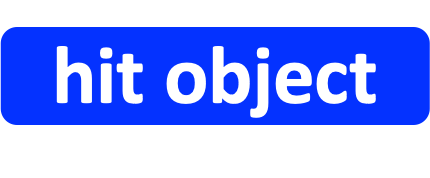}}]), it means the component is optional. When multiple components are in parenthesis and separated by a vertical bar (\eg (\raisebox{-0.16cm}{\includegraphics[height=0.05\linewidth]{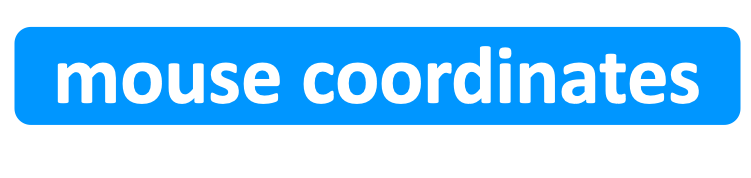}} | \raisebox{-0.16cm}{\includegraphics[height=0.05\linewidth]{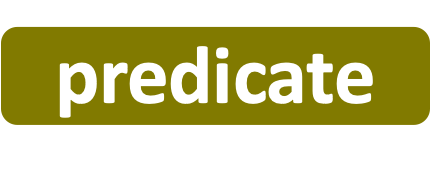}}) ),  it means only one of the components is needed. It should be noted that the techniques presented here are not meant to be exhaustive. 
\subsection{Connects visual objects and changes in visual properties}
\subsubsection{Select: Visually Differentiate Objects from Surroundings}
\technique{Point Select (S) \& Multi-Select (S)}
\noindent(\raisebox{-0.16cm}{\includegraphics[height=0.05\linewidth]{figures/hitobject.png}} | \raisebox{-0.16cm}{\includegraphics[height=0.05\linewidth]{figures/predicate.png}})
\raisebox{-0.16cm}{\includegraphics[height=0.05\linewidth]{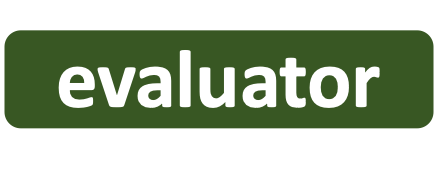}}
\raisebox{-0.16cm}{\includegraphics[height=0.05\linewidth]{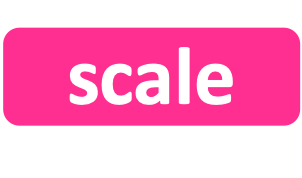}}
\\
In point selection, a visual object is marked and visually differentiated from its surroundings. For example, in Figure \ref{fig:pointselect} \cite{alexander_bank_nodate}, clicking on a glyph representing the date 12/21/2017 changes its stroke color and width, and makes the other glyphs opaque. To support point select, evaluate every target visual object to determine if it is the hit object; alternatively, create a predicate (\eg \textit{data} $=$ 12/21/2017), and evaluate the targets against the predicate.  The evaluation result is then used to encode the visual channels. Multi-select works in a similar way, except that the predicate operand is a list of data values, and the operator is \textit{in}. For example, in Figure \ref{fig:multiselect} \cite{parlapiano_how_2014}, six selected polylines representing six industries are highlighted in green, while the other polylines turn gray and opaque. Point select and multi-select can be triggered by mouse events such as clicking (Figure \ref{fig:pointselect}) and scrolling (Figure \ref{fig:multiselect}), as well as UI events such as a button click.

\begin{figure}[htbp]
\centering
\begin{subfigure}[b]{0.6\linewidth}
     \centering
     \includegraphics[width=\linewidth]{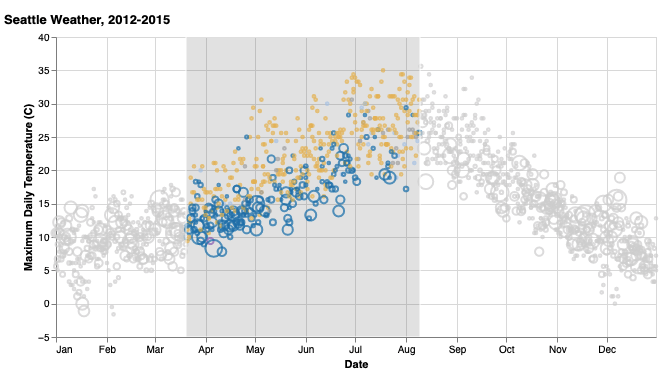}
     \caption{Select a date range 
     }
     \label{fig:rangeselect}
 \end{subfigure}
 \begin{subfigure}[b]{0.36\linewidth}
     \centering
     \includegraphics[width=\linewidth]{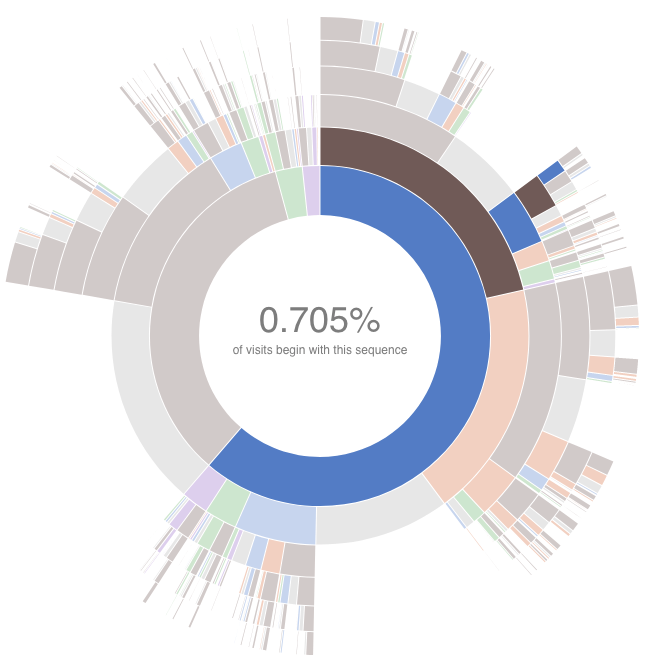}
     \caption{Select a tree branch 
     }
     \label{fig:generalizedselect}
 \end{subfigure}
 \vspace{-3mm}
\caption{Range Select and Generalized Select}
\label{fig:select2}
\end{figure}

\technique{Range Select (S)}
\noindent(\raisebox{-0.16cm}{\includegraphics[height=0.05\linewidth]{figures/mouse.png}} | \raisebox{-0.16cm}{\includegraphics[height=0.05\linewidth]{figures/predicate.png}})
\raisebox{-0.16cm}{\includegraphics[height=0.05\linewidth]{figures/evaluator.png}}
\raisebox{-0.16cm}{\includegraphics[height=0.05\linewidth]{figures/scale.png}}
\\
To range select visual objects along the \textit{x} or \textit{y} dimension (\eg Figure \ref{fig:rangeselect}) \cite{vega-lite_seattle_nodate}, either (1) evaluate every target visual object to determine if it intersects with pixel range defined by the mouse event (click+drag); or (2) identify the data field encoding the position (\eg \textit{date} in Figure \ref{fig:rangeselect}), convert the pixel range into a predicate (\eg \textit{date} $\in$ [3/20, 8/6]), and evaluate if each target's date value satisfies the predicate. The evaluation result is used to encode the visual channel values, where targets failing the evaluation turn gray.

\technique{Generalized Select (S, M)} 
\raisebox{-0.16cm}{\includegraphics[height=0.05\linewidth]{figures/hitobject.png}}
\raisebox{-0.16cm}{\includegraphics[height=0.05\linewidth]{figures/predicate.png}}
\raisebox{-0.16cm}{\includegraphics[height=0.05\linewidth]{figures/evaluator.png}}
\raisebox{-0.16cm}{\includegraphics[height=0.05\linewidth]{figures/scale.png}}
\\
Generalized selection \cite{heer_generalized_2008} expands the hit object to a broader set of visual objects. For example, hovering over an arc representing a webpage click selects the arc as well as arcs representing the previous clicks in a sunburst chart (Figure \ref{fig:generalizedselect}) \cite{kerry_sequences_nodate}. To support generalized selection, create a predicate based on the hit object (\eg \textit{clickID} $\in$ [\textit{sequence from root to ``c5''}]), then evaluate each target against the predicate, finally, encode the visual channel (\ie opacity) values according to the evaluation result.

\technique{Linked Select (M)} 
(\raisebox{-0.16cm}{\includegraphics[height=0.05\linewidth]{figures/mouse.png}} | \raisebox{-0.16cm}{\includegraphics[height=0.05\linewidth]{figures/hitobject.png}})
\raisebox{-0.16cm}{\includegraphics[height=0.05\linewidth]{figures/predicate.png}}
\raisebox{-0.16cm}{\includegraphics[height=0.05\linewidth]{figures/evaluator.png}}
\raisebox{-0.16cm}{\includegraphics[height=0.05\linewidth]{figures/scale.png}}
\\
To enable linked selection (\eg Figure \ref{fig:coordinate}b) across multiple scenes, create a predicate based on either the hit object (if present) or the mouse coordinates, evaluate the target visual objects 
against the predicate, and encode the visual channel values according to the evaluation result.

\technique{Deselect (S, M)}
[\raisebox{-0.16cm}{\includegraphics[height=0.05\linewidth]{figures/hitobject.png}}]
\raisebox{-0.16cm}{\includegraphics[height=0.05\linewidth]{figures/predicate.png}}
\raisebox{-0.16cm}{\includegraphics[height=0.05\linewidth]{figures/evaluator.png}}
\raisebox{-0.16cm}{\includegraphics[height=0.05\linewidth]{figures/scale.png}}
\\
Two forms of deselection exist: users can click a selected object to remove it from selection, or they can click  an empty spot in the background to clear selection. Both forms can be implemented as updating an existing predicate. For the former, remove the hit object ID from the predicate value operand; for the latter, set the predicate value operand to null. The evaluator is then invoked and the targets' visual channels are encoded according to the evaluation results.

\subsubsection{Annotate: Show Auxiliary Visual Objects}
\label{sec:annotate}
\begin{figure}[hbp]
\centering
\begin{subfigure}[b]{0.45\linewidth}
     \centering
     \includegraphics[width=\linewidth]{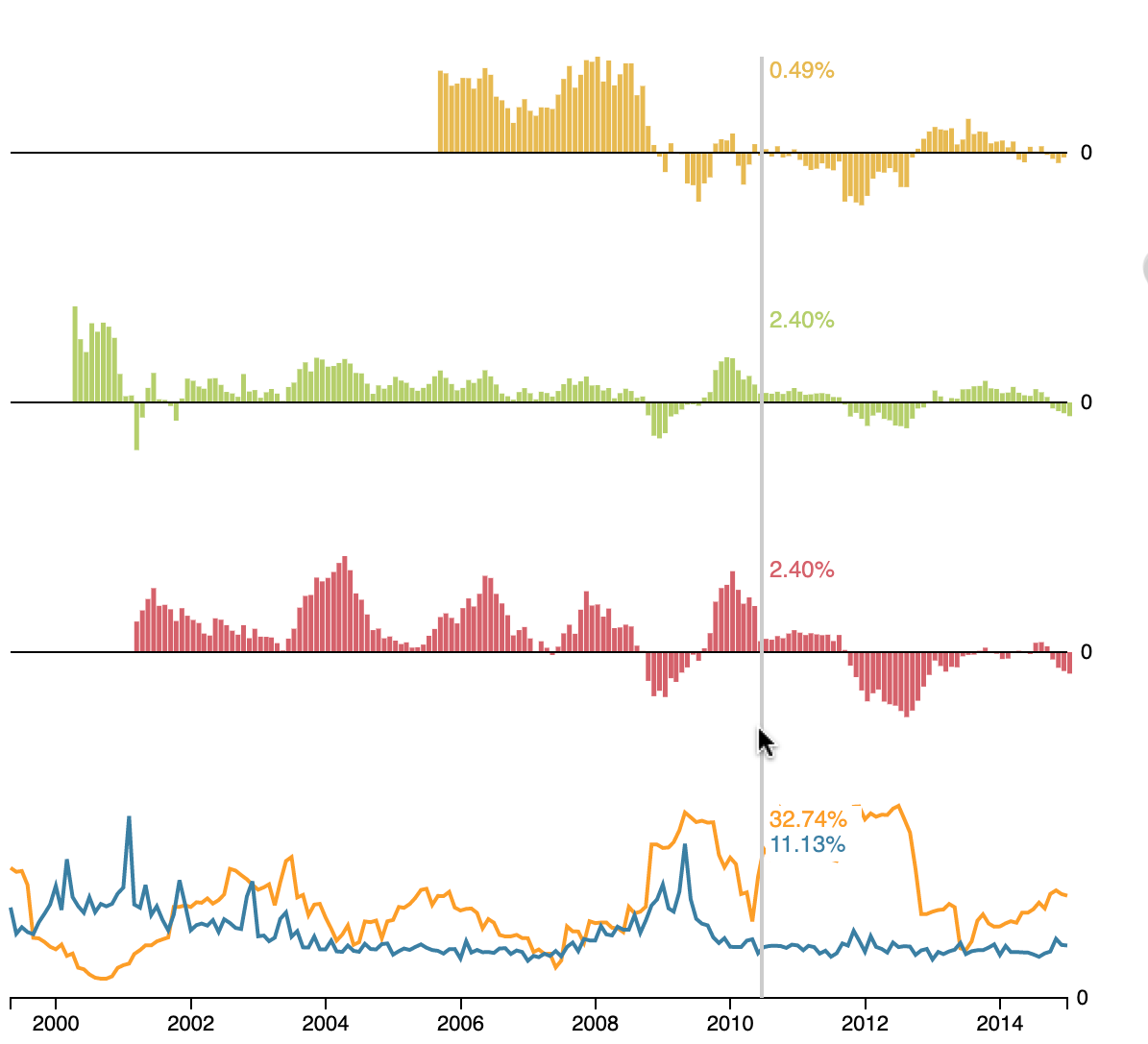}
     \caption{Indicate selected date 
     }
     \label{fig:showreferenceobj1}
 \end{subfigure}
 \begin{subfigure}[b]{0.52\linewidth}
     \centering
    \includegraphics[width=\linewidth]{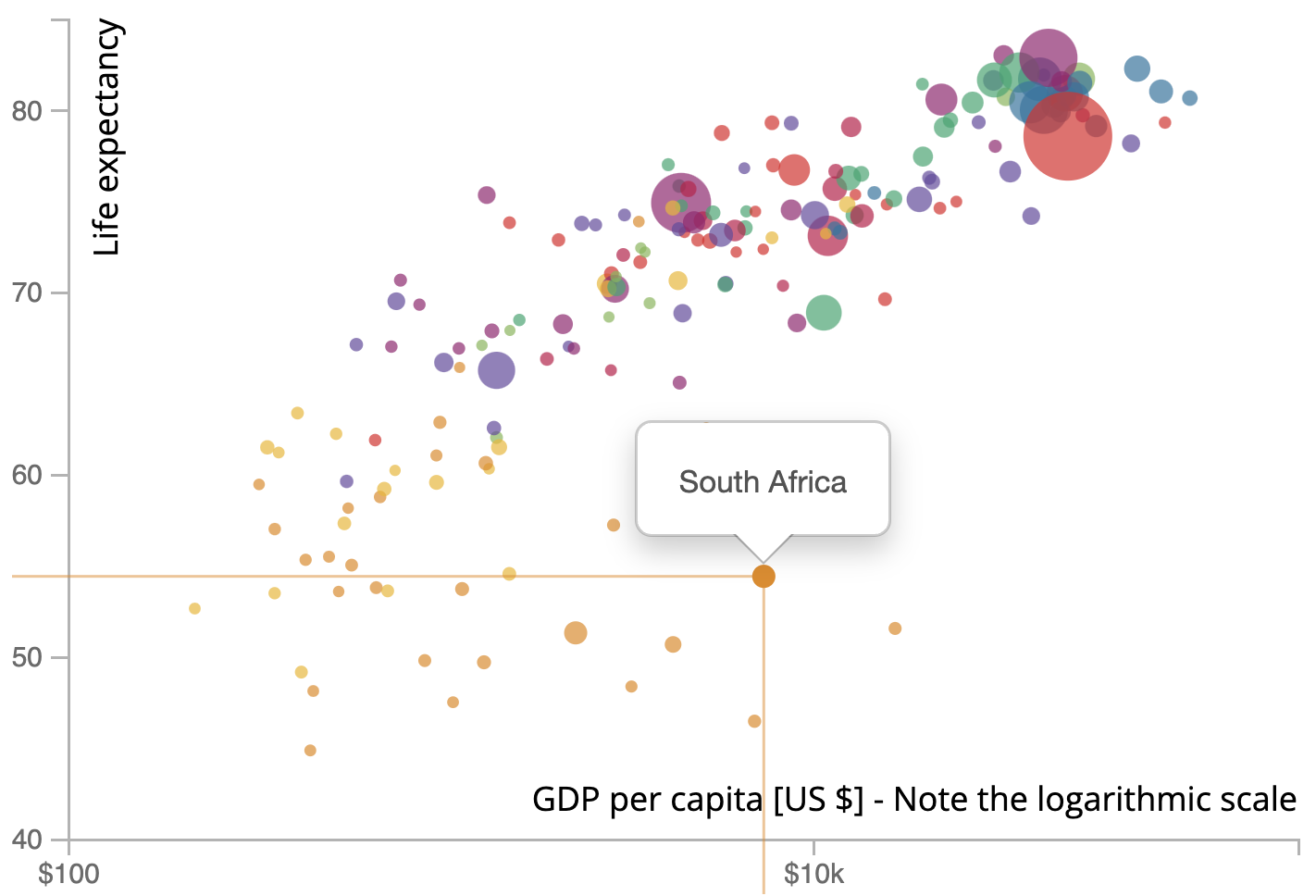}
     \caption{Facilitate value reading}
     \label{fig:showreferenceobj2}
 \end{subfigure}
\vspace{-3mm}
\caption{Annotate: Show Reference Lines}
\label{fig:annotate}
\end{figure}

\technique{Show/Hide Reference Lines (S)}
\noindent(\raisebox{-0.16cm}{\includegraphics[height=0.05\linewidth]{figures/hitobject.png}} | \raisebox{-0.16cm}{\includegraphics[height=0.05\linewidth]{figures/predicate.png}})
[\raisebox{-0.16cm}{\includegraphics[height=0.05\linewidth]{figures/mouse.png}}]
\\
A reference line is often shown to indicate the predicate value used for selection, 
or to facilitate visual estimation of data values. For example, in Figure \ref{fig:showreferenceobj1} \cite{wolf_alpha_nodate}, hovering on a set of related charts sharing the same x-axis shows a vertical line to help users keep track of the location of their action and the currently selected date. 
In another example (Figure \ref{fig:showreferenceobj2}) \cite{scatterplot_with_voronoi_life_nodate}, hovering over a circle shows two orthogonal orange lines, intended to help users read the x and y values more accurately.  
In our analysis, we assume that these reference lines have been created, and interaction primarily changes their visual channels such as visibility and position. To enable updating reference object, different components are involved depending on the visual channels and chart design. For example, in Figure \ref{fig:showreferenceobj1}, the vertical line's position is based on mouse x-coordinate, and its visibility depends on whether a valid date predicate operand is found based on mouse x-coordinate; in Figure \ref{fig:showreferenceobj2}, the lines' positions are based on mouse coordinates, and their visibility depends on whether a hit object is found. 

\technique{Show/Hide Tooltip Container (S)}

\noindent(\raisebox{-0.16cm}{\includegraphics[height=0.05\linewidth]{figures/hitobject.png}} | \raisebox{-0.16cm}{\includegraphics[height=0.05\linewidth]{figures/predicate.png}})
[\raisebox{-0.16cm}{\includegraphics[height=0.05\linewidth]{figures/mouse.png}}]

We define tooltips broadly to include visual objects that appear 
when hovering over a glyph or scene to provide contextual details about data items of interest. Tooltips thus include both the balloon element with text content ``South Afria'' in Figure \ref{fig:showreferenceobj2} as well as the colored text labels in Figure \ref{fig:showreferenceobj1} showing percentage values. To show a tooltip, typically a hit object is involved, and the position of the tooltip depends either on the mouse coordinates or the position of the hit object (Figure \ref{fig:showreferenceobj2}). 
A tooltip can also be shown without any hit object, if a predicate is present (Figure \ref{fig:showreferenceobj1}). 
Note that for the \textit{annotate} intent, we only focus on the \textit{display} of auxiliary visual objects, not their contents. Updating content is addressed in the Details on Demand technique under the \textit{filter} intent category, which we will discuss in the next subsection.  

\subsubsection{Reconfigure: Change the Spatial Arrangement of  Existing Objects}

\technique{Reposition (S, M)} (\raisebox{-0.16cm}{\includegraphics[height=0.05\linewidth]{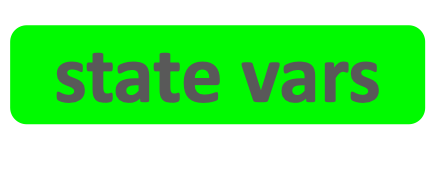}}  | \raisebox{-0.16cm}{\includegraphics[height=0.05\linewidth]{figures/mouse.png}})
In scrollytelling visualizations, it is essential to control the pace of the narrative and dynamically update the positions of visual objects. 
In many cases, the new position is determined by the distance scrolled by mouse; in other cases, a target's new position can be predefined and stored in a state variable. The second case can also be done by mouse click on a UI such as breadcrumb.

\technique{Sort (S)} \raisebox{-0.16cm}{\includegraphics[height=0.05\linewidth]{figures/statevars.png}} \raisebox{-0.16cm}{\includegraphics[height=0.05\linewidth]{figures/evaluator.png}}

Sorting changes the order of visual objects based on field values. 
To enable sorting, a field reference \textit{orderBy} needs to be defined, and its value can be updated by UI controls such as a dropdown menu.
An order evaluator then computes the new list of object indices. 

\technique{Organize Views (M)} \raisebox{-0.16cm}{\includegraphics[height=0.05\linewidth]{figures/mouse.png}} \raisebox{-0.16cm}{\includegraphics[height=0.05\linewidth]{figures/evaluator.png}}
Organize views is a technique to change the positions and sizes of scenes in a composite visualization. For example, in a multiple coordinated view design, 
by clicking on a view and dragging them to a new position, the view layout can be adjusted based on a layout evaluator embedded in the visualization system.

\subsection{Connects visual objects and changes in visible area}
\subsubsection{Steer: Direct Camera View} 
\begin{figure}[htbp]
 \centering
 \includegraphics[width=\linewidth]{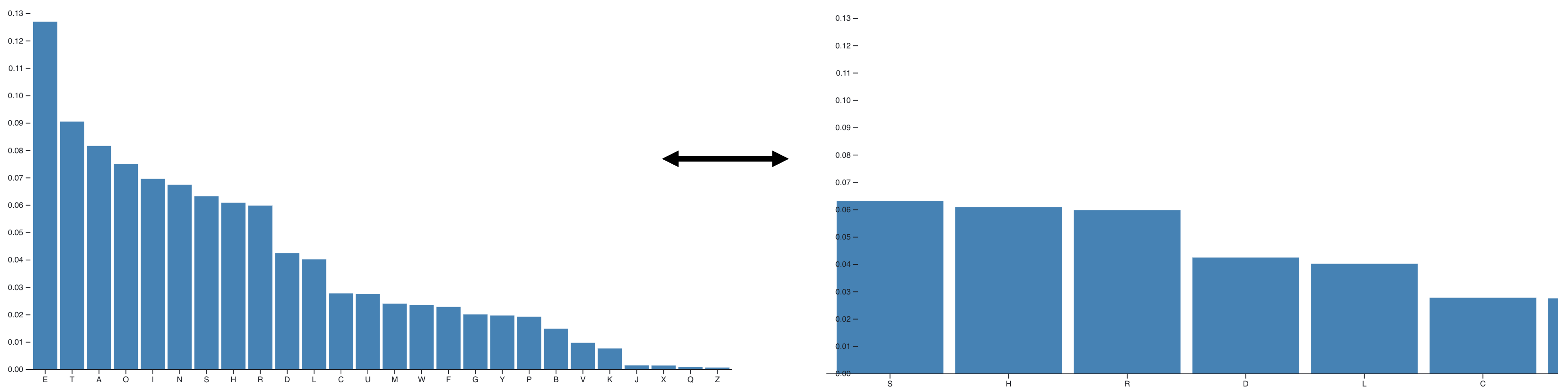}
 \caption{Zoom in and out on a bar chart
 }
\label{fig:zoom+pan}
\vspace{-5mm}
\end{figure}

towards Certain Objects
\technique{Geometric Zoom (S)}
\raisebox{-0.16cm}{\includegraphics[height=0.05\linewidth]{figures/mouse.png}}  \raisebox{-0.16cm}{\includegraphics[height=0.05\linewidth]{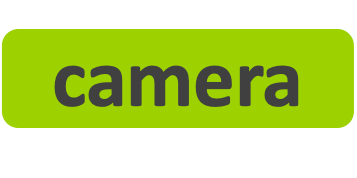}}
\\
Geometric zoom updates the visible area of a scene by reconfiguring camera focus point and zoom level based on mouse events. Figure \ref{fig:zoom+pan} \cite{bostock_zoomable_nodate} shows a zoomable bar chart. Users double click anywhere in the chart or pinch through mouse wheel to perform zooming. To enable geometric zoom, the coordinates of the input mouse event typically become the new focus point. 
The zoom level is usually updated by fixed increments based on the amount of pinching or the number of double clicking. 



\technique{Pan (S)} 
\raisebox{-0.16cm}{\includegraphics[height=0.05\linewidth]{figures/mouse.png}} \raisebox{-0.16cm}{\includegraphics[height=0.05\linewidth]{figures/camera.png}}
\\
Unlike geometric zoom which can update both a camera's focus point and zoom level, panning only changes the focus point based on mouse coordinates. Clicking \& Dragging is the most common mouse event to perform panning (\eg Figure \ref{fig:zoom+pan}) 
Both geometric zooming and panning take place at the scene level without a hit visual object, and everything inside the scene becomes a target. 

\begin{figure}[!b]
\centering
\begin{subfigure}[b]{1\linewidth}
     \centering
     \includegraphics[width=\linewidth]{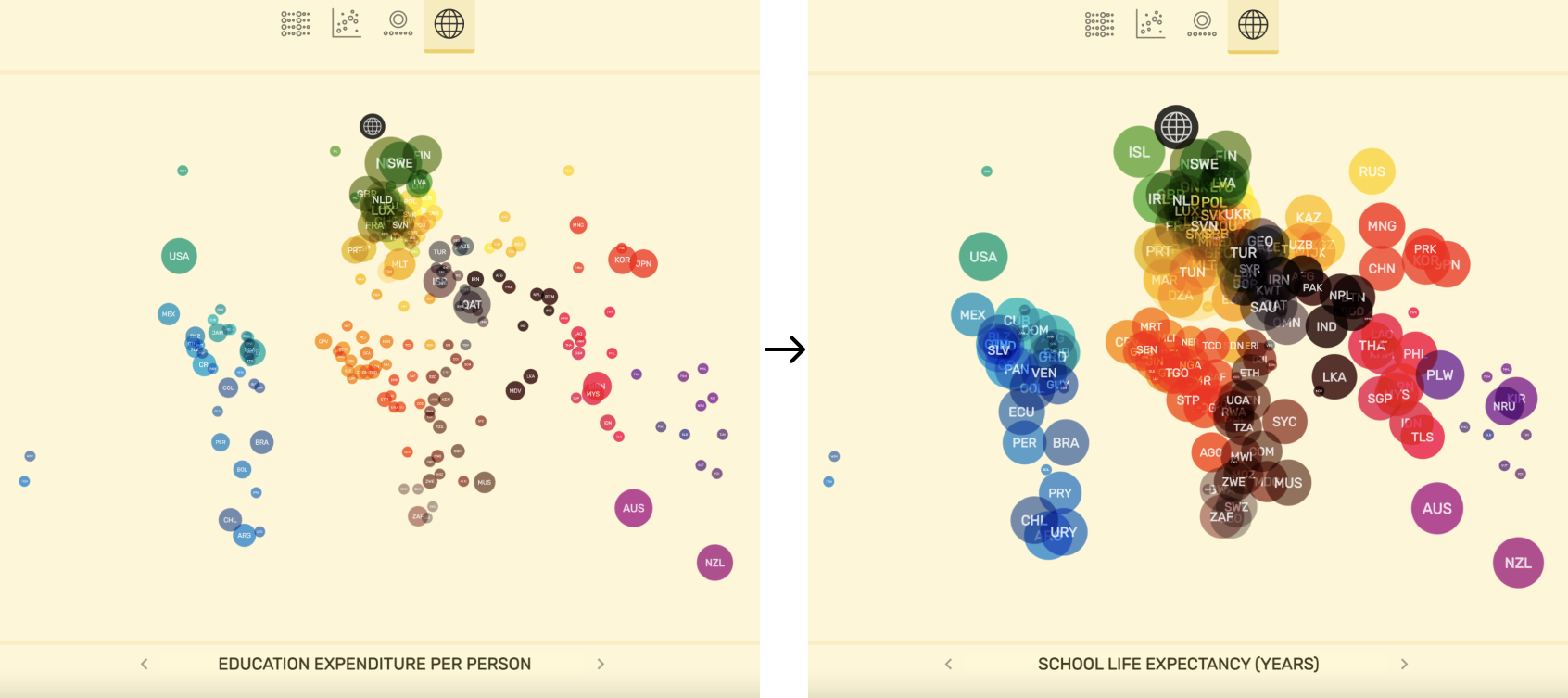}
     \caption{Change field in size encoding   \vspace{2mm}
     }
     \label{fig:changeFieldEncoding}
 \end{subfigure}

 \begin{subfigure}[b]{1\linewidth}
     \centering
     \includegraphics[width=\linewidth]{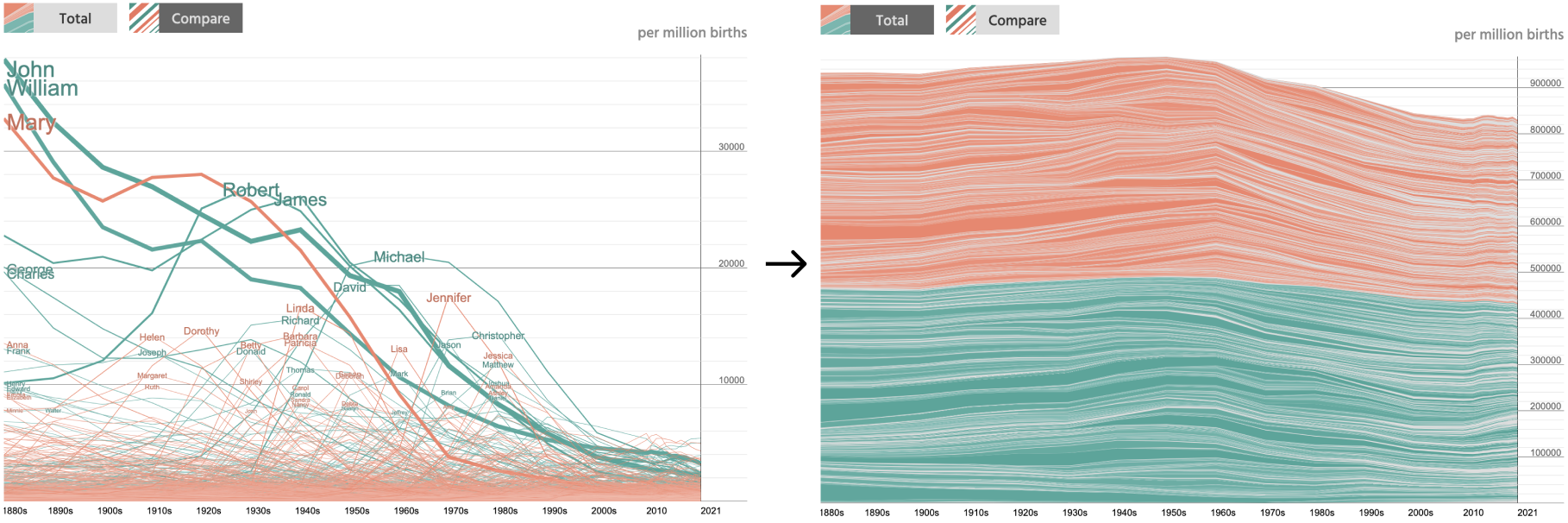}
     \caption{Change from multi-line chart to stacked area chart 
     }
     \label{fig:changeChartType}
 \end{subfigure}
\vspace{-4mm}
\caption{Encode: Change Field and Chart Type}
\label{fig:change_encoding}
\end{figure}

\technique{Toggle between Different Views (M)} (\raisebox{-0.16cm}{\includegraphics[height=0.05\linewidth]{figures/camera.png}} | \raisebox{-0.16cm}{\includegraphics[height=0.05\linewidth]{figures/statevars.png}})
In tab-based user interfaces, it is common to display a view at a time by clicking on the UI control. For example, in GOV | DNA, by clicking on one of the four buttons at the top (\eg Figure \ref{fig:changeFieldEncoding}), the currently visible view is updated. The views are usually predefined, each containing unique definitions of components, such as marks, encodings, and even predicates (to control which data items are visualized). Depending on how the views are defined, this technique can be realized by updating either a state variable or the camera. In Figure \ref{fig:changeFieldEncoding}, the views are significantly different, and it is cleaner to treat them as independent scene definitions. Toggling between them thus requires the definition of a state variable \textit{currentScene}. In examples such as the Star Mapper \cite{interactive_esa_nodate}, the toggling is more easily described by updating the camera's position and focus point in a three-dimensional scene. 

\begin{figure}[htbp]
 \centering
 \includegraphics[width=\linewidth]{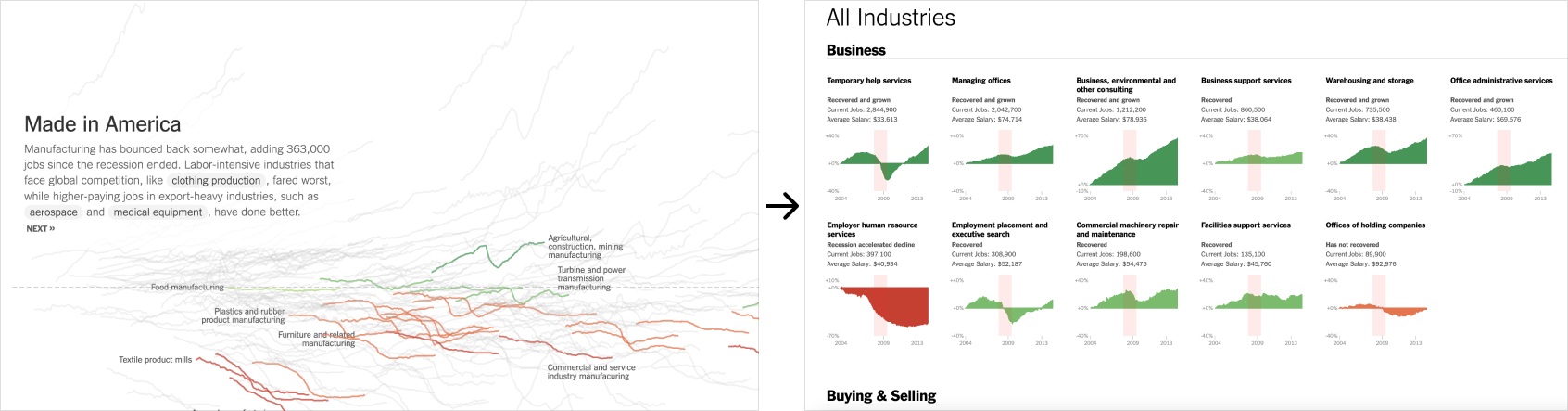}
 \caption{Navigating to the next section in a narrative visualization}
\label{fig:scrollscene}
\end{figure}

\technique{Navigate to a Scene or Section (M)}
\raisebox{-0.16cm}{\includegraphics[height=0.05\linewidth]{figures/statevars.png}} [\raisebox{-0.16cm}{\includegraphics[height=0.05\linewidth]{figures/mouse.png}}]
\\
Similar to Toggle between Different Views, this technique changes the current scene or section; however, this technique is primarily used in narrative visualizations. For example, based on the mouse coordinates scrolled (\eg Figure \ref{fig:scrollscene}) or the UI control clicked on  (\eg breadcrumb in Figure \ref{fig:multiselect}), the component reference \textit{currentScene} updates, changing the currently visible view. 

\vspace{-3mm}
\subsection{Connects visual objects and changes in visual mapping}
\subsubsection{Encode: Change the Visual Representation of Existing Data Items}
\technique{Change Field in Encoding (S)} 
\raisebox{-0.16cm}{\includegraphics[height=0.05\linewidth]{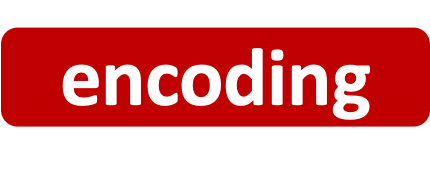}} \raisebox{-0.16cm}{\includegraphics[height=0.05\linewidth]{figures/scale.png}}
\\
In Figure \ref{fig:changeFieldEncoding}, users can click on the left/right arrow buttons to choose a data field to encode the size of circles. Assuming that an encoder has already been created to map data values to a given visual channel properties, to enable changing field in encoding, we do not need to update the target's data, we simply update the data field in the encoding component definition. 


\technique{Change Chart Type (S)} 
\raisebox{-0.16cm}{\includegraphics[height=0.05\linewidth]{figures/statevars.png}}
\raisebox{-0.16cm}{\includegraphics[height=0.05\linewidth]{figures/encoding.png}} 
\raisebox{-0.16cm}{\includegraphics[height=0.05\linewidth]{figures/scale.png}}\\
In Figure \ref{fig:changeChartType} \cite{laura_namegrapher_nodate}, clicking on the buttons (``Total'' or ``Compare'') toggles the visualization between a multi-line graph and a stacked area chart, showing the same data items. Multiple aspects of visual encoding are changed: mark type, visual channels and layout. To implement changing chart type, define state variables referencing the chart type definition and associated encodings, and create encoders that handle the mapping between data values to the corresponding marks and channels.  


\vspace{4mm}
\subsubsection{Enter/Exit}
\technique{Click to Add Data Points (S)}
\raisebox{-0.16cm}{\includegraphics[height=0.05\linewidth]{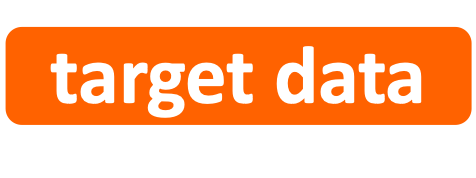}}\\
In our example collection, we only encountered one case that introduces new data items through interaction. In a D3.js demo \cite{mouseEvent_Will}, clicking the  background will add data points which are shown as new circle marks at the clicked locations. This example shows how interaction can modify the data being visualized.

\subsection{Connects visual objects and changes in data}
\subsubsection{Filter}
\technique{Dynamic Queries (S, M)}
\raisebox{-0.16cm}{\includegraphics[height=0.05\linewidth]{figures/predicate.png}}
\noindent(\raisebox{-0.16cm}{\includegraphics[height=0.05\linewidth]{figures/evaluator.png}}
\raisebox{-0.16cm}{\includegraphics[height=0.05\linewidth]{figures/scale.png}} | \raisebox{-0.16cm}{\includegraphics[height=0.05\linewidth]{figures/targetdata.png}})\\
Dynamic Queries \cite{ahlberg_dynamic_1992,shneiderman_dynamic_1994} allow users to control which data items should be visualized through interaction with user interface elements such as sliders, checkboxes, and manipulable widgets \cite{hochheiser_dynamic_2004,wattenberg_sketching_2001} that embody filtering criteria. 
To support dynamic queries, define a predicate for each querying control, and update predicate operands based on user interaction. The target visual objects can then be updated in two ways. First, we can evaluate the targets against a predicate evaluator, and encode the targets' visibility accordingly, but this approach will not work for visualizations that require re-computation of layouts (\eg stacked area chart in Figure \ref{fig:changeChartType}). Second, we can treat the scene as the target and update its data based on the predicate. The scene can then be updated with the same encoding mechanisms. 



\technique{Details-on-Demand (S)}
\raisebox{-0.16cm}{\includegraphics[height=0.05\linewidth]{figures/hitobject.png}}
\raisebox{-0.16cm}{\includegraphics[height=0.05\linewidth]{figures/predicate.png}}
\raisebox{-0.16cm}{\includegraphics[height=0.05\linewidth]{figures/targetdata.png}}
\\
Figure \ref{fig:tooltip} is a canonical example of showing details on demand: when hovering over a mark, the tooltip content updates to show detailed information about the item of interest. We assume that the text elements in the tooltip and the mapping between the text elements and corresponding data fields have been created. Updating the content thus requires retrieving the relevant data items, which can be categorized as \textit{filter}. To support details-on-demand, obtain a predicate from the hit object, and set the tooltip's data as the data items satisfying the predicate. In our analysis, handling the visibility and position of the tooltip is a separate operation, which falls under the \textit{annotate} category, as discussed in Section \ref{sec:annotate}.


\technique{Cross-filter (M)}
[\raisebox{-0.16cm}{\includegraphics[height=0.05\linewidth]{figures/hitobject.png}}]
\raisebox{-0.16cm}{\includegraphics[height=0.05\linewidth]{figures/predicate.png}}
\noindent(\raisebox{-0.16cm}{\includegraphics[height=0.05\linewidth]{figures/evaluator.png}}
\raisebox{-0.16cm}{\includegraphics[height=0.05\linewidth]{figures/scale.png}} | \raisebox{-0.16cm}{\includegraphics[height=0.05\linewidth]{figures/targetdata.png}})
\\
Cross-filter occurs when user interaction in one scene filters the data represented by visual objects in other scenes (Figure \ref{fig:cross-filter}). 
The implementation of cross-filter is similar to that of dynamic queries: create a predicate based on input event, then either evaluate the target objects against the predicate and encode the evaluation results accordingly, or change the target scene's data. 


\subsubsection{Abstract/Elaborate: Change the Levels of Abstraction}
\begin{figure}[htbp]
 \centering
 \includegraphics[width=\linewidth]{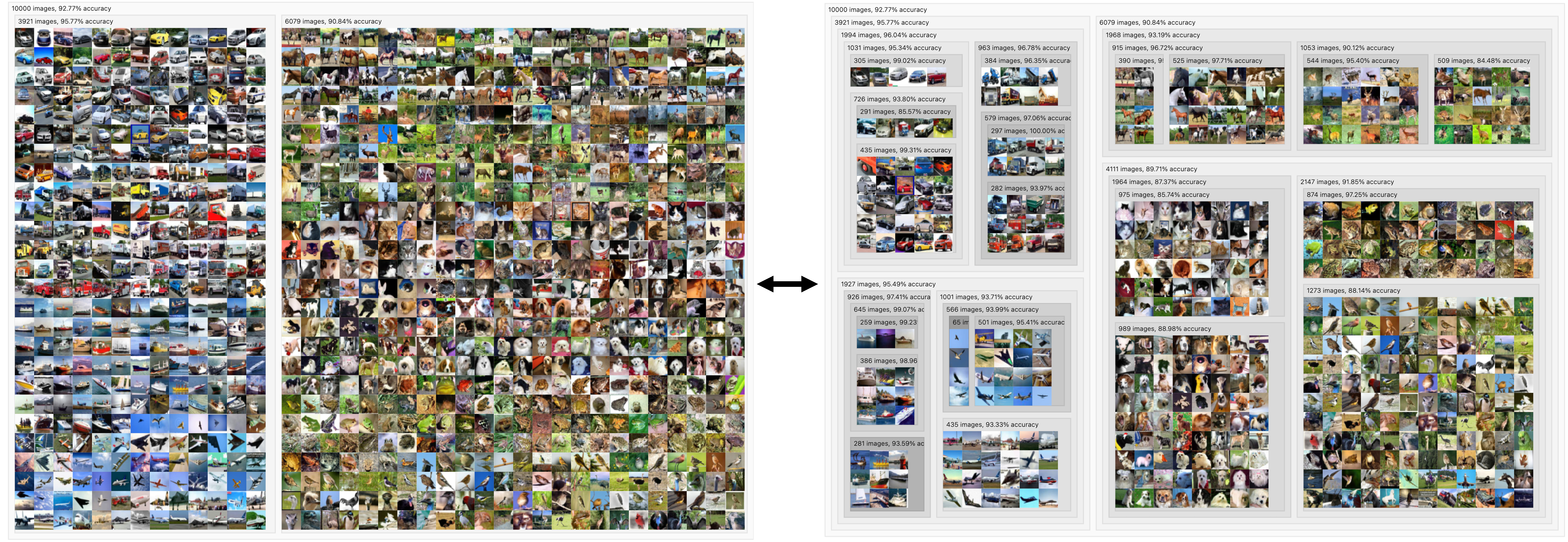}
 \caption{Change the level of abstraction in a hierarchy}
\label{fig:dendromap}
\end{figure}

\technique{Move Up/Down in a Hierarchy} \raisebox{-0.16cm}{\includegraphics[height=0.05\linewidth]{figures/statevars.png}} \raisebox{-0.16cm}{\includegraphics[height=0.05\linewidth]{figures/targetdata.png}} \\
Given a hierarchical dataset, users can choose whether to display the entire structure (including all the leaf nodes and nodes at the intermediate levels), or only nodes above a certain level. For example, DendroMap \cite{bertucci_dendromap_2022} displays hierarchical clusters of image datasets. Users can increase or decrease the number of visible clusters (Figure \ref{fig:dendromap}) through a slider. The interaction sets the level of abstraction in the underlying hierarchy. To enable moving up and down a hierarchy, create a state variable recording the desired level of abstraction, and update the target data whenever the variable value changes. 

\technique{Drill Down/Roll Up} \raisebox{-0.16cm}{\includegraphics[height=0.05\linewidth]{figures/statevars.png}} \raisebox{-0.16cm}{\includegraphics[height=0.05\linewidth]{figures/targetdata.png}} \\
Levels of abstraction can be changed on multivariate dataset too. In Polestar \cite{moritz_polestar_nodate}, Polaris \cite{stolte_polaris_2002} and Tableau \cite{tableau_software_tableau_nodate}, users can drill down data by dragging additional dimensions to the row or column shelf, which will partition existing data displays into a finer grained aggregation.  To enable drill-down and roll-up, create state variables recording the current dimensions and measure, and update the target data (based on pre-computed data cubes or real-time querying) whenever the variable values change. 


\subsubsection{Derive: Compute and Show New Field Values from Data Items Attached to Existing Objects}

\technique{Recompute Field with New Baseline (S)} 
\raisebox{-0.16cm}{\includegraphics[height=0.05\linewidth]{figures/statevars.png}}
\raisebox{-0.16cm}{\includegraphics[height=0.05\linewidth]{figures/targetdata.png}}

Many visualizations show numeric measures computed against a baseline. Figure \ref{fig:linked-select} shows the temporal trend of a currency's (\eg Swiss franc) Big Mac Index against a base currency (\eg US dollar). Choosing a different baseline will generate a new set of index values that are encoded as the height of the lollipops. To enable re-computation against a dynamic baseline, define a state variable representing the baseline, and change the target scene's data. 


\begin{figure}[htbp]
 \centering
 \includegraphics[width=0.8\linewidth]{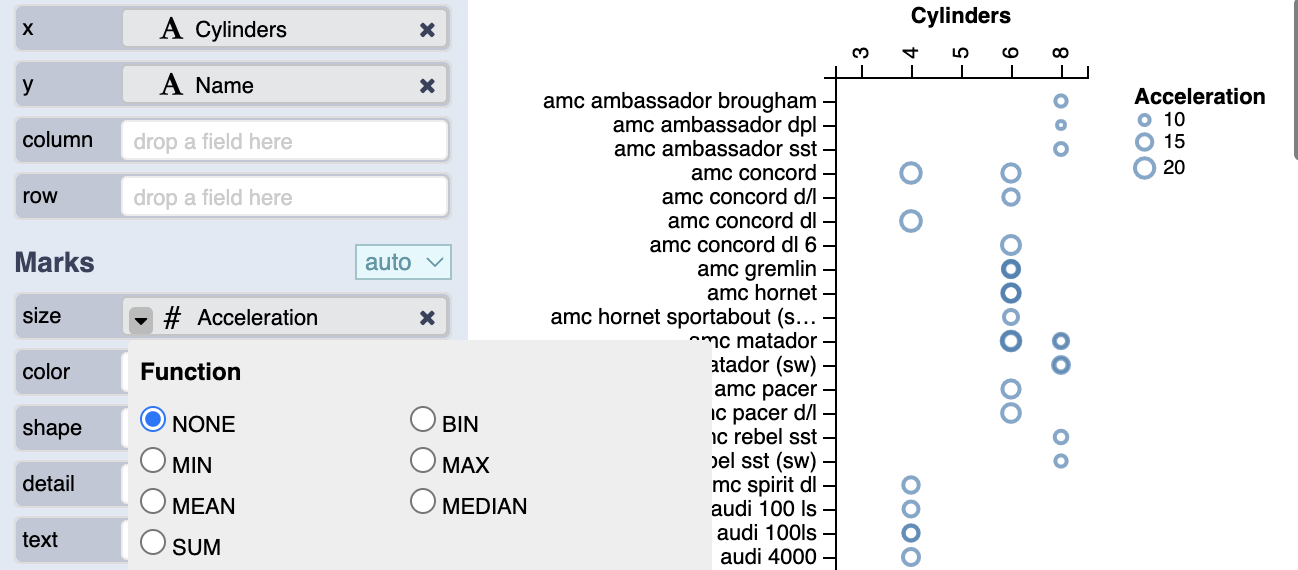}
 \caption{Change aggregator function to compute values}
\label{fig:aggregator}
\end{figure}

\technique{Change Aggregator (S)} 
\raisebox{-0.16cm}{\includegraphics[height=0.05\linewidth]{figures/statevars.png}}
\raisebox{-0.16cm}{\includegraphics[height=0.05\linewidth]{figures/targetdata.png}}
\\
In visualizations where a mark represents multiple data items, it is possible to change how the values for a field are aggregated. For example, Polestar \cite{moritz_polestar_nodate}, Polaris \cite{stolte_polaris_2002} and Tableau \cite{tableau_software_tableau_nodate} allow users to choose from a list of aggregation functions (Figure \ref{fig:aggregator}). To support changing aggregator, define a state variable representing the aggregator, and update the target scene's data whenever the aggregator is changed. 

\begin{table*}[!b]\fontfamily{ptm}\selectfont\small\smaller
\centering
\begin{tabular} {p{1.8cm}p{1.2cm}p{2cm}p{1.2cm}p{1.6cm}p{1.5cm}p{1.7cm}p{2.5cm}}
\arrayrulecolor{lightgray}
\toprule
& User Intent & Technique & Event & Listener & Hit Object & Target & Internal Components\\
\midrule
Add Magnet &  enter & add mark & choose field & dropdown menu & none & new magnet & target data\\ 
Move Magnet & reconfigure & reposition & click + drag  & canvas & magnet & all dust particles & component references, target evaluator\\ 
\bottomrule
\end{tabular}
\caption{Dust and Magnet}
\label{tbl:dnm}

\begin{tabular} {p{1.8cm}p{1.2cm}p{2cm}p{1.2cm}p{1.6cm}p{1.5cm}p{1.7cm}p{2.5cm}}
\arrayrulecolor{lightgray}
\toprule
& User Intent & Technique & Event & Listener & Hit Object & Target & Internal Components\\
\midrule
Add Matrix & enter & add collection & choose set & set list & none & new matrix & target data\\
Move Matrix & select & point select & click + drag & canvas & matrix & all matrices & predicate, target evaluator, evaluation scale \\ 
Compare Matrices & abstract & move up a hierarchy & drag end & canvas & matrix & all matrices & target data\\ 
Change Operator & derive & change aggregator & click & canvas & operator mark  & matrix & state variable (logical operator), target data\\ 
\bottomrule
\end{tabular}
\caption{OnSet}
\label{tbl:onset}
\end{table*}

\subsection{Multiple User Intents}
We also encountered many interaction designs demonstrating more than one intent. We present two such examples below. 

\begin{figure}[t]
\centering
\begin{subfigure}[b]{0.49\linewidth}
     \centering
     \includegraphics[width=\linewidth]{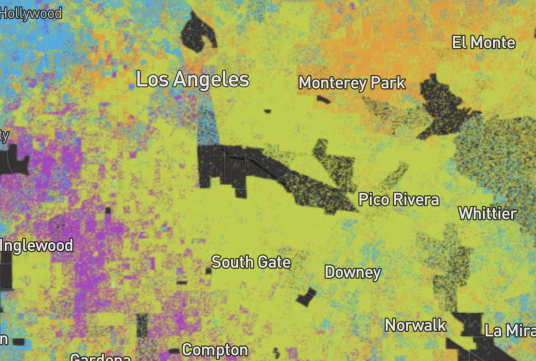}
     \caption{Semantic Zoom}
     \label{fig:sementiczoom1}
\end{subfigure}
\begin{subfigure}[b]{0.49\linewidth}
     \centering
     \includegraphics[width=\linewidth]{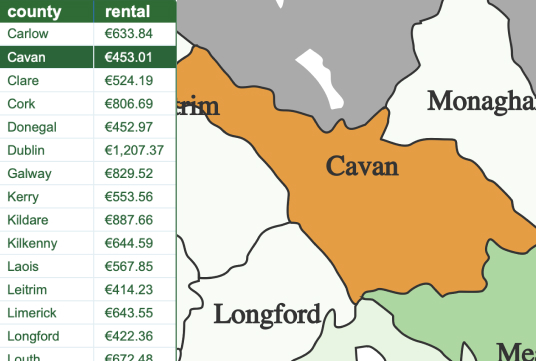}
     \caption{Direct Walk}
     \label{fig:directwalk2}
\end{subfigure}
\caption{Multiple Intents: Semantic Zoom and Direct Walk}
\label{fig:filter1}
\end{figure}

\technique{Semantic Zoom: Steer + Abstract/Elaborate (S)}
\raisebox{-0.16cm}{\includegraphics[height=0.05\linewidth]{figures/camera.png}} \raisebox{-0.16cm}{\includegraphics[height=0.05\linewidth]{figures/statevars.png}} \component{targetdata.png}
\\
Unlike geometric zoom, semantic zoom has dual intents: steer and abstract/elaborate. When zooming in the heatmap in Figure \ref{fig:sementiczoom1} \cite{heavyai_1990_nodate}, not only the camera zoom level changes, the bin size of the heatmap also updates to show more granular information. 

\technique{Direct Walk: Select + Steer (M)} (\raisebox{-0.16cm}{\includegraphics[height=0.05\linewidth]{figures/hitobject.png}} | \raisebox{-0.16cm}{\includegraphics[height=0.05\linewidth]{figures/predicate.png}})  \raisebox{-0.16cm}{\includegraphics[height=0.05\linewidth]{figures/camera.png}} \raisebox{-0.16cm}{\includegraphics[height=0.05\linewidth]{figures/evaluator.png}} \raisebox{-0.16cm}{\includegraphics[height=0.05\linewidth]{figures/scale.png}}
\\
Direct walk ``smoothly move[s] the viewing
focus from one position in information structure to another'' \cite{yi_toward_2007}. In Figure \ref{fig:directwalk2} \cite{kerr_d3_nodate}, clicking on a table row both selects a county map polygon and performs a geometric zoom with the county as the focus point. 


%% file: sections/8._discussion_and_conclusion.tex


\section{Discussion: Descriptive, Evaluative, and Generative Powers and Completeness}
We evaluate the descriptive, evaluative, and generative powers \cite{beaudouin-lafon_designing_2004} of our framework, and discuss the completeness of the taxonomies. 

\vspace{-2mm}
\subsection{Descriptive Power} 
\label{sec:descriptive_power}
To understand how well our framework can describe complex and bespoke interaction designs, we choose Dust and Magnet (DnM) \cite{soo_yi_dust_2005} and OnSet \cite{sadana_onset_2014} as two case studies, based on the following considerations: (1) they use customized static visual representations instead of standard chart types, and (2) the interactive behaviors deviate from the representative techniques included in our analysis.
 
\bpstart{Dust and Magnet} DnM allows users to interactively explore a multivariate dataset. Users can add one or more fields as magnets. All the data items are shown as iron particles. When moving a magnet representing a field, the dust particles are attracted to the magnet, where the particles with high values for the given field will move closer to the magnet. Table \ref{tbl:dnm} analyzes two primary actions in DnM using our framework. For adding a magnet, The authoring intent is to link the user input to the relevant system components, thereby generating a new magnet. The user intent is to introduce a new field. Choosing an option from the dropdown menu adds a data field, which is visualized as a magnet. For moving a magnet, the authoring intent is to link the user input to the relevant system components, thereby changing positions of the magnet. The user intent is to reconfigure. Consequently, we need a target evaluator that takes all the data items as input, and computes the new positions for each item. To perform this computation, the target evaluator keeps track of all the references to the existing magnets to read their positions and fields.

\bpstart{OnSet} Developed to visualize sets, OnSet uses matrices to represent a single set, logical operations on multiple sets, or nested logical operations on multiple sets. Each data item is assigned a fixed position in the matrices, and its presence or absence in a set is shown using color. The appearance of a matrix varies accordingly to the data it represents (i.e., a single set or composite sets). Table \ref{tbl:onset} analyzes four primary actions supported in Onset, encompassing three authoring intents and four user intents.
Both adding a matrix and comparing matrices modify the target data: the former adds a new set (a tree with only one node) which is visualized as a matrix; the latter constructs a tree where the leaf nodes are individual sets and their ancestors are results of applying logical operators on the children sets. Note that OnSet has defined encodings and scales for determining the visual representations of a tree depending on its structure and content, and these rules remain independent from the interactions.
When moving a matrix (the hit object), a predicate is used by a target evaluator to determine if the bounding box of each matrix in the visualization intersects with the hit object, and an evaluation scale assigns visual properties to each matrix based on the evaluation result. 
Finally, to support toggling between ``AND'' and ``OR'' operators on a composite matrix, the program needs to keep a state variable for the operator and update the matrix's data whenever the operator changes.





\vspace{-2mm}
\subsection{Evaluative Power}
Our framework provides a theoretical lens and vocabulary to evaluate existing and new languages and tools for composing interactive visualizations. The evaluation can be conducted across three aspects: levels of abstraction, expressiveness, flexibility.

\bpstart{Levels of abstraction} The first aspect, levels of abstraction is to examine where a system primarily operates at. Each system encompasses different levels of abstraction, which defines the range of interaction authoring in the system. For example, Interactive+ is to augment an existing static web-based visualization with interactivity. This tool operates primarily at the component level: given a visualization realized in the HTML Document Object Model \cite{whatwg_dom_nodate}, Interaction+ extracts component-level information including the visual objects and their attributes as well as the distribution of visual objects over each visual attribute. Interaction+’s user interface does not provide explicit abstractions of concepts at the technique and user intent level. As an example of a system that covers multiple abstraction levels, consider Lyra 2, which grants access to two levels of abstraction. At the component level, author's demonstration is primarily in the form of choosing visual marks in a static visualization generated by Vega or Vega-Lite, and constructing query objects that are translated to predicates. An interaction inspector also provides component-level information about recommended design (\eg mouse event, visual attributes); at the technique level, the suggested interaction techniques are shown as selectable widgets with visual thumbnails in the inspector. Lyra 2 does not provide explicit visual representations or interaction mechanisms at the user intent level.

\bpstart{Expressiveness} The second aspect, expressiveness, evaluates the range of interactivity against the categories at the identified levels of abstraction. For instance, Interaction+ is tool-agnostic, allowing users to add interactions to a static web-based visualization without accessing the underlying data. Consequently, this tool supports some of the techniques such as generalized selection without a predicate component involving data updates. It does this by defining filtering criteria based on visual attributes alone. The exclusion of data engagement restricts expressiveness by not supporting some of the data-related techniques (e.g., Click to add data point, Change field in encoding, Recompute Field with New Baseline and more). Even for techniques such as generalized selection, if a data field is not visually encoded, it would not be possible to select based on the field in Interaction+. With Vega and Vega-Lite as the underlying grammars for static visualizations, Lyra 2 has access to a much more complete set of components than Interaction+ (e.g., encodings and data). It is thus more expressive, supporting techniques such as generalized selection involving data projection. However, some of richer techniques are unavailable (e.g., Navigate to Previous/Next Scene or Section, Organize view, and Reposition). As interactive visualizations have become highly expressive, they demand advanced techniques, such as scrolly-telling and drag-and-drop dashboards. Thus, with Lyra 2, how to compose such interactions remains unclear.

\bpstart{Flexibility} The third aspect, flexibility, refers to the extent to which tools or languages can be easily adjusted or expanded to accommodate users' specific requirements. Take the example of Interaction+ as we previously discussed regarding its limitation in terms of expressiveness, the tool also exhibits inflexibility. This is primarily due to the restricted techniques stemming from omitting data when authoring interactions, the unclear definitions of their six categories of user intents, and the absence of UI abstractions for techniques. Such limitations make the tool difficult to flexibly operate across multi-levels of abstraction. Overall, Interaction+ is primarily created for use cases where interaction centers around visual objects and attributes, and it may not be sufficient to provide support for either fine-grained interaction design activities involving data and state variables, or concise specification of high-level authoring intents and techniques. Meanwhile, Lyra 2 exhibits a considerable amount of flexibility. By providing visual widgets representing the suggested interaction design, authors can think and interact with the concepts at the technique level. They also can easily inspect and change lower-level components such as mouse events and visual channels. Users cannot specify interaction design at the intent level since the interface provides no abstractions at that level.

In this section, we showcase the evaluative power of our framework by comparing two examples with different levels of abstraction, expressiveness, and flexibility. Our framework occupies a middle ground between template-based tools (e.g., Tableau) and low-level visualization libraries (e.g., D3.js). To encompass all levels of abstraction represented in our framework, we selected two non-template-based authoring systems that incorporate multiple techniques and operate at a component level of abstraction. Based on our understanding, there are limited systems that fulfill these criteria. Therefore, we plan to develop an interaction authoring grammar to showcase our framework's evaluative power further.

\vspace{-4mm}
\subsection{Generative Power}
Our analysis identifies components associated with each technique, which can inform the implementation of interactivity in visualization. It also has the potential to help create new languages and systems for interaction authoring. We demonstrate the generative power of our framework by proposing new ideas and guidelines.

First of all, both Interaction+ and Lyra 2 focus on creating interactivity for marks, and mostly overlook the other types of visual objects such as collection and scene. By including these visual objects in the scope, techniques such as OnSet, scrollytelling and dynamic toggling are possible. Second, it is important that an expressive interaction authoring language or tool needs to understand and provide internal abstractions for various representation and interaction components. It would also be helpful to expose these components in the interface through manipulable visual representations by adopting, for example, a visual programming paradigm \cite{kuhail_characterizing_2021}. Finally, a powerful authoring tool can allow users to start at any of the three levels of abstraction, depending on their expertise and tasks, and intelligently provide assistance through appropriate abstractions. The introduction mentioned one example where an intent can be translated into techniques and low-level implementations. We can imagine a different scenario where an author has composed an interaction technique using certain components, and wants to see how the implementation can be generalized to other techniques and even intents. 

\vspace{-2mm}
\subsection{Completeness}
When selecting real-world examples, we sought to increase the diversity in terms of chart type and source (Section \ref{sec:method}). We do not claim, however, that the categories at each level of our framework are comprehensive and complete, capturing all possible interaction designs. First of all, we can only include representative techniques for each intent category. Many interaction designs, including the two in Section \ref{sec:descriptive_power}, bear multiple intents. The list of techniques is thus certainly not exhaustive. For the intent and component categories, it is not our primary goal to ensure the taxonomies are complete. As demonstrated in scientific fields like biology where classifications of life forms are constantly updated, we anticipate that our taxonomies will evolve with new entries and revisions as novel interaction designs are created.

%% file: sections/9._conclusion.tex
\vspace{-2mm}

\section{Conclusion}
We present a unified framework of the task space of interaction authoring in data visualization at three levels: authoring intent, technique, and component. We identify conceptual categories at each level, and discuss how these categories can be linked for a holistic understanding of interaction design. This analysis represents a step towards a systematic theoretical framework for interactive visualization design.

%% file: main.bib
@article{zong_lyra_2020,
	title = {Lyra 2: {Designing} interactive visualizations by demonstration},
	volume = {27},
	shorttitle = {Lyra 2},
	number = {2},
	journal = {IEEE Transactions on Visualization and Computer Graphics},
	author = {Zong, Jonathan and Barnwal, Dhiraj and Neogy, Rupayan and Satyanarayan, Arvind},
	year = {2020},
	note = {Publisher: IEEE},
	pages = {304--314},
}

@article{satyanarayan_vega-lite_2016,
	title = {Vega-lite: {A} grammar of interactive graphics},
	volume = {23},
	shorttitle = {Vega-lite},
	number = {1},
	journal = {IEEE transactions on visualization and computer graphics},
	author = {Satyanarayan, Arvind and Moritz, Dominik and Wongsuphasawat, Kanit and Heer, Jeffrey},
	year = {2016},
	note = {Publisher: IEEE},
	pages = {341--350},
}

@article{morth_scrollyvis_2022,
	title = {{ScrollyVis}: {Interactive} visual authoring of guided dynamic narratives for scientific scrollytelling},
	shorttitle = {{ScrollyVis}},
	journal = {IEEE Transactions on Visualization and Computer Graphics},
	author = {Mörth, Eric and Bruckner, Stefan and Smit, Noeska N.},
	year = {2022},
	note = {Publisher: IEEE},
}

@inproceedings{lu_automatic_2021,
	title = {Automatic {Generation} of {Unit} {Visualization}-based {Scrollytelling} for {Impromptu} {Data} {Facts} {Delivery}},
	booktitle = {2021 {IEEE} 14th {Pacific} {Visualization} {Symposium} ({PacificVis})},
	publisher = {IEEE},
	author = {Lu, Junhua and Chen, Wei and Ye, Hui and Wang, Jie and Mei, Honghui and Gu, Yuhui and Wu, Yingcai and Zhang, Xiaolong Luke and Ma, Kwan-Liu},
	year = {2021},
	pages = {21--30},
}

@inproceedings{wattenberg_sketching_2001,
	title = {Sketching a graph to query a time-series database},
	booktitle = {{CHI}'01 {Extended} {Abstracts} on {Human} factors in {Computing} {Systems}},
	author = {Wattenberg, Martin},
	year = {2001},
	pages = {381--382},
}

@article{latif_kori_2022,
	title = {Kori: {Interactive} {Synthesis} of {Text} and {Charts} in {Data} {Documents}},
	volume = {28},
	issn = {1941-0506},
	shorttitle = {Kori},
	doi = {10.1109/TVCG.2021.3114802},
	number = {1},
	journal = {IEEE Transactions on Visualization and Computer Graphics},
	author = {Latif, Shahid and Zhou, Zheng and Kim, Yoon and Beck, Fabian and Kim, Nam Wook},
	month = jan,
	year = {2022},
	note = {Conference Name: IEEE Transactions on Visualization and Computer Graphics},
	pages = {184--194},
}

@inproceedings{sultanum_leveraging_2021,
	title = {Leveraging text-chart links to support authoring of data-driven articles with vizflow},
	booktitle = {Proceedings of the 2021 {CHI} {Conference} on {Human} {Factors} in {Computing} {Systems}},
	author = {Sultanum, Nicole and Chevalier, Fanny and Bylinskii, Zoya and Liu, Zhicheng},
	year = {2021},
	pages = {1--17},
}

@article{heer_interactive_2012,
	title = {Interactive dynamics for visual analysis},
	volume = {55},
	issn = {0001-0782},
	url = {https://doi.org/10.1145/2133806.2133821},
	doi = {10.1145/2133806.2133821},
	number = {4},
	urldate = {2022-04-26},
	journal = {Communications of the ACM},
	author = {Heer, Jeffrey and Shneiderman, Ben},
	month = apr,
	year = {2012},
	pages = {45--54},
}

@article{yi_toward_2007,
	title = {Toward a {Deeper} {Understanding} of the {Role} of {Interaction} in {Information} {Visualization}},
	volume = {13},
	issn = {1941-0506},
	doi = {10.1109/TVCG.2007.70515},
	number = {6},
	journal = {IEEE Transactions on Visualization and Computer Graphics},
	author = {Yi, Ji Soo and Kang, Youn ah and Stasko, John and Jacko, J.A.},
	month = nov,
	year = {2007},
	note = {Conference Name: IEEE Transactions on Visualization and Computer Graphics},
	pages = {1224--1231},
}

@article{satyanarayan_critical_2019,
	title = {Critical reflections on visualization authoring systems},
	volume = {26},
	number = {1},
	journal = {IEEE transactions on visualization and computer graphics},
	author = {Satyanarayan, Arvind and Lee, Bongshin and Ren, Donghao and Heer, Jeffrey and Stasko, John and Thompson, John and Brehmer, Matthew and Liu, Zhicheng},
	year = {2019},
	note = {Publisher: IEEE},
	pages = {461--471},
}

@inproceedings{zhicheng_liu_atlas_2021,
	title = {Atlas: {Grammar}-based {Procedural} {Generation} of {Data} {Visualizations}},
	booktitle = {2021 {IEEE} {Visualization} {Conference} ({VIS})},
	author = {{Zhicheng Liu} and Chen, Chen and Morales, Francisco and Zhao, Yishan},
	year = {2021},
}

@article{satyanarayan_lyra_2014,
	title = {Lyra: {An} {Interactive} {Visualization} {Design} {Environment}},
	volume = {33},
	copyright = {© 2014 The Author(s) Computer Graphics Forum © 2014 The Eurographics Association and John Wiley \& Sons Ltd. Published by John Wiley \& Sons Ltd.},
	issn = {1467-8659},
	shorttitle = {Lyra},
	url = {https://onlinelibrary.wiley.com/doi/abs/10.1111/cgf.12391},
	doi = {https://doi.org/10.1111/cgf.12391},
	language = {en},
	number = {3},
	urldate = {2021-02-13},
	journal = {Computer Graphics Forum},
	author = {Satyanarayan, Arvind and Heer, Jeffrey},
	year = {2014},
	pages = {351--360},
}

@article{ren_charticulator_2018,
	title = {Charticulator: {Interactive} construction of bespoke chart layouts},
	volume = {25},
	shorttitle = {Charticulator},
	number = {1},
	journal = {IEEE transactions on visualization and computer graphics},
	author = {Ren, Donghao and Lee, Bongshin and Brehmer, Matthew},
	year = {2018},
	note = {Publisher: IEEE},
	pages = {789--799},
}

@article{tsandilas_structgraphics_2020,
	title = {{StructGraphics}: {Flexible} {Visualization} {Design} through {Data}-{Agnostic} and {Reusable} {Graphical} {Structures}},
	shorttitle = {{StructGraphics}},
	journal = {IEEE Transactions on Visualization and Computer Graphics},
	author = {Tsandilas, Theophanis},
	year = {2020},
	note = {Publisher: IEEE},
}

@article{buja_interactive_1996,
	title = {Interactive high-dimensional data visualization},
	volume = {5},
	number = {1},
	journal = {Journal of computational and graphical statistics},
	author = {Buja, Andreas and Cook, Dianne and Swayne, Deborah F.},
	year = {1996},
	note = {Publisher: Taylor \& Francis},
	pages = {78--99},
}

@incollection{shneiderman_eyes_2003,
	address = {San Francisco},
	series = {Interactive {Technologies}},
	title = {The {Eyes} {Have} {It}: {A} {Task} by {Data} {Type} {Taxonomy} for {Information} {Visualizations}},
	isbn = {978-1-55860-915-0},
	shorttitle = {The {Eyes} {Have} {It}},
	url = {https://www.sciencedirect.com/science/article/pii/B9781558609150500469},
	language = {en},
	urldate = {2021-03-01},
	booktitle = {The {Craft} of {Information} {Visualization}},
	publisher = {Morgan Kaufmann},
	author = {Shneiderman, Ben},
	editor = {Bederson, BENJAMIN B. and Shneiderman, BEN},
	month = jan,
	year = {2003},
	doi = {10.1016/B978-155860915-0/50046-9},
	pages = {364--371},
}

@online{whatwg_dom_nodate,
	title = {{DOM} Standard},
	author = {{WHATWG}},
	year = {n.d.},
	url = {https://dom.spec.whatwg.org/},
	note = {Available at: \url{https://dom.spec.whatwg.org/} (Accessed: 13 June 2022)},
	urldate = {2022-06-13},
}

@inproceedings{liu_data_2018,
	title = {Data {Illustrator}: {Augmenting} vector design tools with lazy data binding for expressive visualization authoring},
	shorttitle = {Data {Illustrator}},
	booktitle = {Proceedings of the 2018 {CHI} {Conference} on {Human} {Factors} in {Computing} {Systems}},
	author = {Liu, Zhicheng and Thompson, John and Wilson, Alan and Dontcheva, Mira and Delorey, James and Grigg, Sam and Kerr, Bernard and Stasko, John},
	year = {2018},
	pages = {1--13},
}

@article{bostock_d3_2011,
	title = {D3: data-driven documents},
	volume = {17},
	number = {12},
	journal = {IEEE transactions on visualization and computer graphics},
	author = {Bostock, Michael and Ogievetsky, Vadim and Heer, Jeffrey},
	year = {2011},
	note = {Publisher: IEEE},
	pages = {2301--2309},
}

@inproceedings{beaudouin-lafon_instrumental_2000,
	title = {Instrumental interaction: an interaction model for designing post-{WIMP} user interfaces},
	shorttitle = {Instrumental interaction},
	booktitle = {Proceedings of the {SIGCHI} conference on {Human} factors in computing systems},
	author = {Beaudouin-Lafon, Michel},
	year = {2000},
	pages = {446--453},
}

@inproceedings{beaudouin-lafon_designing_2004,
	title = {Designing interaction, not interfaces},
	booktitle = {Proceedings of the working conference on {Advanced} visual interfaces},
	author = {Beaudouin-Lafon, Michel},
	year = {2004},
	pages = {15--22},
}

@inproceedings{heer_generalized_2008,
	title = {Generalized selection via interactive query relaxation},
	booktitle = {Proceedings of the {SIGCHI} {Conference} on {Human} {Factors} in {Computing} {Systems}},
	author = {Heer, Jeffrey and Agrawala, Maneesh and Willett, Wesley},
	year = {2008},
	pages = {959--968},
}

@book{foley_computer_1996,
	title = {Computer graphics: principles and practice},
	volume = {12110},
	shorttitle = {Computer graphics},
	publisher = {Addison-Wesley Professional},
	author = {Foley, James D. and Van, Foley Dan and Van Dam, Andries and Feiner, Steven K. and Hughes, John F.},
	year = {1996},
}

@inproceedings{thompson_understanding_2020,
	title = {Understanding the design space and authoring paradigms for animated data graphics},
	volume = {39},
	booktitle = {Computer {Graphics} {Forum}},
	publisher = {Wiley Online Library},
	author = {Thompson, John and Liu, Zhicheng and Li, Wilmot and Stasko, John},
	year = {2020},
	note = {Issue: 3},
	pages = {207--218},
}

@inproceedings{dix_starting_1998,
	address = {L'Aquila, Italy},
	title = {Starting simple: adding value to static visualisation through simple interaction},
	shorttitle = {Starting simple},
	url = {http://portal.acm.org/citation.cfm?doid=948496.948514},
	doi = {10.1145/948496.948514},
	language = {en},
	urldate = {2022-11-23},
	booktitle = {Proceedings of the working conference on {Advanced} visual interfaces  - {AVI} '98},
	publisher = {ACM Press},
	author = {Dix, Alan and Ellis, Geoffrey},
	year = {1998},
	pages = {124},
}

@article{stolte_polaris_2002,
	title = {Polaris: {A} system for query, analysis, and visualization of multidimensional relational databases},
	volume = {8},
	shorttitle = {Polaris},
	number = {1},
	journal = {IEEE Transactions on Visualization and Computer Graphics},
	author = {Stolte, Chris and Tang, Diane and Hanrahan, Pat},
	year = {2002},
	note = {Publisher: IEEE},
	pages = {52--65},
}

@inproceedings{ahlberg_dynamic_1992,
	title = {Dynamic queries for information exploration: {An} implementation and evaluation},
	shorttitle = {Dynamic queries for information exploration},
	booktitle = {Proceedings of the {SIGCHI} conference on {Human} factors in computing systems},
	author = {Ahlberg, Christopher and Williamson, Christopher and Shneiderman, Ben},
	year = {1992},
	pages = {619--626},
}

@article{shneiderman_dynamic_1994,
	title = {Dynamic queries for visual information seeking},
	volume = {11},
	number = {6},
	journal = {IEEE software},
	author = {Shneiderman, Ben},
	year = {1994},
	note = {Publisher: IEEE},
	pages = {70--77},
}

@article{hochheiser_dynamic_2004,
	title = {Dynamic query tools for time series data sets: timebox widgets for interactive exploration},
	volume = {3},
	shorttitle = {Dynamic query tools for time series data sets},
	number = {1},
	journal = {Information Visualization},
	author = {Hochheiser, Harry and Shneiderman, Ben},
	year = {2004},
	note = {Publisher: SAGE Publications Sage UK: London, England},
	pages = {1--18},
}

@online{scatterplot_with_voronoi_life_nodate,
	title = {Life expectancy versus {GDP} per Capita},
	url = {https://bl.ocks.org/anonymous/raw/0ff9728abfc3e86d9b3af6d529f47278/8507af7883fff0d75d70db6bb62849216c10ea45/},
	note = {Available at \url{https://bl.ocks.org/anonymous/raw/0ff9728abfc3e86d9b3af6d529f47278/8507af7883fff0d75d70db6bb62849216c10ea45/}},		
	author = {{Scatterplot with Voronoi}},
	urldate = {2022-11-25},
}

@online{moritz_polestar_nodate,
	title = {Polestar},
	url = {https://vega.github.io/polestar/},
	note = {Available at \url{https://vega.github.io/polestar/}},			
	author = {Moritz, Dominik and Wongsuphasawat, Kanit and Heer, Jeffrey},
	urldate = {2022-11-25},
}

@online{tableau_software_tableau_nodate,
	title = {Tableau: Business Intelligence and Analytics Software},
	url = {https://www.tableau.com/},
	note = {Available at \url{https://www.tableau.com/}},		
	shorttitle = {Tableau},
	titleaddon = {Tableau},
	author = {{Tableau Software}},
	urldate = {2021-06-04},
	langid = {american},
}

@article{bertucci_dendromap_2022,
	title = {{DendroMap}: {Visual} {Exploration} of {Large}-{Scale} {Image} {Datasets} for {Machine} {Learning} with {Treemaps}},
	issn = {1941-0506},
	shorttitle = {{DendroMap}},
	doi = {10.1109/TVCG.2022.3209425},
	journal = {IEEE Transactions on Visualization and Computer Graphics},
	author = {Bertucci, Donald and Hamid, Md Montaser and Anand, Yashwanthi and Ruangrotsakun, Anita and Tabatabai, Delyar and Perez, Melissa and Kahng, Minsuk},
	year = {2022},
	note = {Conference Name: IEEE Transactions on Visualization and Computer Graphics},
	pages = {1--11},
}

@inproceedings{wang_falx_2021,
	title = {Falx: {Synthesis}-{Powered} {Visualization} {Authoring}},
	shorttitle = {Falx},
	booktitle = {Proceedings of the 2021 {CHI} {Conference} on {Human} {Factors} in {Computing} {Systems}},
	author = {Wang, Chenglong and Feng, Yu and Bodik, Rastislav and Dillig, Isil and Cheung, Alvin and Ko, Amy J.},
	year = {2021},
	pages = {1--15},
}

@article{munzner_multi-level_2013,
	title = {A multi-level typology of abstract visualization tasks},
	volume = {19},
	number = {12},
	journal = {IEEE transactions on visualization and computer graphics},
	author = {Munzner, Tamara and Brehmer, Matthew},
	year = {2013},
	note = {Publisher: IEEE},
	pages = {2376--2385},
}

@article{liu_immens_2013,
	title = {{imMens}: {Real}-time {Visual} {Querying} of {Big} {Data}},
	volume = {32},
	issn = {1467-8659},
	shorttitle = {{imMens}},
	url = {https://onlinelibrary.wiley.com/doi/abs/10.1111/cgf.12129},
	doi = {10.1111/cgf.12129},
	language = {en},
	number = {3pt4},
	urldate = {2022-02-08},
	journal = {Computer Graphics Forum},
	author = {Liu, Zhicheng and Jiang, Biye and Heer, Jeffrey},
	year = {2013},
	pages = {421--430},
}

@article{roth_empirically-derived_2013,
	title = {An empirically-derived taxonomy of interaction primitives for interactive cartography and geovisualization},
	volume = {19},
	issn = {1941-0506},
	doi = {10.1109/TVCG.2013.130},
	language = {eng},
	number = {12},
	journal = {IEEE transactions on visualization and computer graphics},
	author = {Roth, Robert E.},
	month = dec,
	year = {2013},
	pmid = {24051802},
	pages = {2356--2365},
}

@article{pike_science_2009,
	title = {The {Science} of {Interaction}},
	volume = {8},
	issn = {1473-8716},
	url = {https://doi.org/10.1057/ivs.2009.22},
	doi = {10.1057/ivs.2009.22},
	language = {en},
	number = {4},
	urldate = {2022-12-01},
	journal = {Information Visualization},
	author = {Pike, William A. and Stasko, John and Chang, Remco and O'Connell, Theresa A.},
	month = jan,
	year = {2009},
	note = {Publisher: SAGE Publications},
	pages = {263--274},
}

@article{engestrom_activity_1999,
	title = {Activity theory and individual and social transformation},
	volume = {19},
	number = {38},
	journal = {Perspectives on activity theory},
	author = {Engeström, Yrjö},
	year = {1999},
	pages = {19--30},
}

@book{brown_perspectives_1999,
	title = {Perspectives on activity theory},
	publisher = {Cambridge university press},
	author = {Brown, John Seely and Heath, Christian and Pea, Roy},
	year = {1999},
}

@article{gotz_characterizing_2009,
	title = {Characterizing users' visual analytic activity for insight provenance},
	volume = {8},
	number = {1},
	journal = {Information Visualization},
	author = {Gotz, David and Zhou, Michelle X.},
	year = {2009},
	note = {Publisher: SAGE Publications Sage UK: London, England},
	pages = {42--55},
}

@article{nardi_activity_1996,
	title = {Activity theory and human-computer interaction},
	volume = {436},
	journal = {Context and consciousness: Activity theory and human-computer interaction},
	author = {Nardi, Bonnie A.},
	year = {1996},
	pages = {7--16},
}

@misc{nardi_context_1998,
	title = {Context and consciousness: {Activity} theory and human-computer interaction},
	shorttitle = {Context and consciousness},
	publisher = {University of Toronto Press},
	author = {Nardi, Bonnie},
	year = {1998},
}

@article{kuutti_activity_1996,
	title = {Activity theory as a potential framework for human-computer interaction research},
	volume = {1744},
	journal = {Context and consciousness: Activity theory and human-computer interaction},
	author = {Kuutti, Kari},
	year = {1996},
}

@online{wolf_alpha_nodate,
	title = {Alpha vs. Correlation/Dispersion},
	url = {https://tylernwolf.com/corrdisp/index.html},
	note = {Available at \url{https://tylernwolf.com/corrdisp/index.html}},		
	author = {Wolf, Tyler},
	urldate = {2022-12-02},
}

@online{the_economist_big_nodate,
	title = {The Big Mac index},
	url = {https://www.economist.com/big-mac-index},
	note = {Available at \url{https://www.economist.com/big-mac-index}},
	titleaddon = {The Economist},
	author = {{The Economist}},
	urldate = {2022-12-02},
	langid = {english},
}

@online{alexander_bank_nodate,
	title = {Bank Holidays},
	author = {Alexander, Waleczek},
	year = {n.d.},
	url = {https://public.tableau.com/views/MakeoverMonday-Week44-BankHolidays/BankHolidays},
	note = {Available at: \url{https://public.tableau.com/views/MakeoverMonday-Week44-BankHolidays/BankHolidays} [Accessed 2 December 2022]},
	urldate = {2022-12-02},
	langid = {english},
}

@article{parlapiano_how_2014,
	chapter = {The Upshot},
	title = {How the {Recession} {Reshaped} the {Economy}, in 255 {Charts}},
	issn = {0362-4331},
	url = {https://www.nytimes.com/interactive/2014/06/05/upshot/how-the-recession-reshaped-the-economy-in-255-charts.html, https://www.nytimes.com/interactive/2014/06/05/upshot/how-the-recession-reshaped-the-economy-in-255-charts.html},
	language = {en-US},
	urldate = {2022-12-02},
	journal = {The New York Times},
	author = {Parlapiano, Alicia and Ashkenas, Jeremy},
	month = jun,
	year = {2014},
	note = {Cad: 1},
}

@online{vega-lite_seattle_nodate,
	title = {Seattle Weather Exploration},
	url = {https://vega.github.io/vega-lite/examples/interactive_seattle_weather.html},
	note = {Available at \url{https://vega.github.io/vega-lite/examples/interactive_seattle_weather.html}},
	titleaddon = {Vega-Lite},
	author = {{Vega-Lite}},
	urldate = {2022-12-02},
}

@online{kerry_sequences_nodate,
	title = {Sequences Sunburst},
	url = {https://observablehq.com/@kerryrodden/sequences-sunburst},
	note = {Available at \url{https://observablehq.com/@kerryrodden/sequences-sunburst}},	
	titleaddon = {Observable},
	author = {Kerry, Rodden},
	urldate = {2022-12-02},
}

@online{bostock_zoomable_nodate,
	title = {Zoomable Bar Chart},
	url = {https://observablehq.com/@d3/zoomable-bar-chart},
	note = {Available at \url{https://observablehq.com/@d3/zoomable-bar-chart}},	
	titleaddon = {Observable},
	author = {Bostock, Mike},
	urldate = {2022-12-02},
}

@online{sudox_govdna_nodate,
	title = {{GOV}{\textbar}{DNA}},
	url = {https://govdna.sudox.nl},
	note = {Available at \url{https://govdna.sudox.nl}},
	author = {{Sudox}},
	urldate = {2022-12-02},
	langid = {english},
}

@online{laura_namegrapher_nodate,
	title = {{NameGrapher}},
	url = {https://namerology.com/baby-name-grapher/},
	note = {Available at \url{https://namerology.com/baby-name-grapher/}},
	titleaddon = {Namerology},
	author = {Laura, Wattenberg},
	urldate = {2022-12-02},
	langid = {american},
}

@online{heavyai_1990_nodate,
	title = {1990 - 2020 {US} Census Dot Density Dashboard},
	url = {https://census2-demo.heavy.ai/omnisci/dashboard/26?tab=-Mmd0tXBrEPn33vRyHK1},
	note = {Available at \url{https://census2-demo.heavy.ai/omnisci/dashboard/26?tab=-Mmd0tXBrEPn33vRyHK1}},	
	author = {{Heavy.ai}},
	urldate = {2022-12-02},
}

@online{kerr_d3_nodate,
	title = {D3 with embedded {SVG} map},
	url = {https://bl.ocks.org/catherinekerr/e345a906f8e2bae8d07dbc79f8f04036},
	note = {Available at \url{https://bl.ocks.org/catherinekerr/e345a906f8e2bae8d07dbc79f8f04036}},	
	author = {Kerr, Catherine},
}

@online{interactive_esa_nodate,
	title = {{ESA} Star Mapper},
	url = {https://sci.esa.int/star_mapper/},
	note = {Available at \url{https://sci.esa.int/star_mapper/}},
	titleaddon = {{ESA} Star Mapper},
	author = {Interactive, {TULP}},
	urldate = {2022-12-02},
}

@online{the_new_york_times_new_nodate,
	title = {The New York Times - Breaking News, {US} News, World News and Videos},
	url = {https://www.nytimes.com},
	note = {Available at \url{https://www.nytimes.com}},
	author = {{The New York Times}},
	urldate = {2022-12-02},
	langid = {english},
}

@online{the_pudding_pudding_nodate,
	title = {The Pudding},
	url = {https://pudding.cool/},
	note = {Available at \url{https://pudding.cool/}},
	titleaddon = {The Pudding},
	author = {{The Pudding}},
	urldate = {2022-12-02},
	langid = {english},
}

@online{quartz_quartz_nodate,
	title = {Quartz},
	url = {https://qz.com/},
	note = {Available at \url{https://qz.com/}},
	titleaddon = {Quartz},
	author = {{Quartz}},
	urldate = {2022-12-02},
	langid = {english},
}

@online{popular_blocks_popular_nodate,
	title = {Popular Blocks},
	url = {https://bl.ocks.org/},
	note = {Available at \url{https://bl.ocks.org/}},	
	author = {{Popular Blocks}},
	urldate = {2022-12-02},
}

@online{tableau_tableau_nodate,
	title = {Tableau},
	url = {https://public.tableau.com/app/discover},
	note = {Available at \url{https://public.tableau.com/app/discover}},
	titleaddon = {Tableau Public},
	author = {{Tableau}},
	urldate = {2022-12-02},
	langid = {english},
}

@online{datasketch_datasketch_nodate,
	title = {Datasketch},
	url = {https://www.datasketch.co/},
	note = {Available at \url{https://www.datasketch.co/}},
	author = {{Datasketch}},
	urldate = {2022-12-02},
}

@online{gramener_gramener_nodate,
	title = {Gramener},
	url = {https://gramener.com/},
	note = {Available at \url{https://gramener.com/}},
	author = {{Gramener}},
	urldate = {2022-12-02},
	langid = {american},
}

@online{column_five_media_column_nodate,
	title = {Column Five Media},
	url = {https://www.columnfivemedia.com/},
	note = {Available at \url{https://www.columnfivemedia.com/}},
	titleaddon = {Column Five},
	author = {{Column Five Media}},
	urldate = {2022-12-02},
}

@online{mdn_web_docs_viewbox_nodate,
	title = {{viewBox} - {SVG}: Scalable Vector Graphics},
	url = {https://developer.mozilla.org/en-US/docs/Web/SVG/Attribute/viewBox},
	note = {Available at \url{https://developer.mozilla.org/en-US/docs/Web/SVG/Attribute/viewBox}},
	shorttitle = {{viewBox} - {SVG}},
	author = {{MDN Web Docs}},
	urldate = {2022-12-02},
	langid = {american},
}

@online{sciencedirect_engineering_topics_perspective_nodate,
	title = {Perspective Camera - an overview {\textbar} {ScienceDirect} Topics},
	url = {https://www.sciencedirect.com/topics/engineering/perspective-camera},
	note = {Available at \url{https://www.sciencedirect.com/topics/engineering/perspective-camera}},	
	author = {{ScienceDirect Engineering Topics}},
	note = {Available at \url{https://www.sciencedirect.com/topics/engineering/perspective-camera}},
	urldate = {2022-12-02},
}

@article{kuhail_characterizing_2021,
	title = {Characterizing {Visual} {Programming} {Approaches} for {End}-{User} {Developers}: {A} {Systematic} {Review}},
	volume = {9},
	issn = {2169-3536},
	shorttitle = {Characterizing {Visual} {Programming} {Approaches} for {End}-{User} {Developers}},
	doi = {10.1109/ACCESS.2021.3051043},
	journal = {IEEE Access},
	author = {Kuhail, Mohammad Amin and Farooq, Shahbano and Hammad, Rawad and Bahja, Mohammed},
	year = {2021},
	note = {Conference Name: IEEE Access},
	pages = {14181--14202},
}

@article{sadana_onset_2014,
	title = {Onset: {A} visualization technique for large-scale binary set data},
	volume = {20},
	number = {12},
	journal = {IEEE transactions on visualization and computer graphics},
	author = {Sadana, Ramik and Major, Timothy and Dove, Alistair and Stasko, John},
	year = {2014},
	note = {ISBN: 1077-2626
Publisher: IEEE},
	pages = {1993--2002},
}

@article{soo_yi_dust_2005,
	title = {Dust \& magnet: multivariate information visualization using a magnet metaphor},
	volume = {4},
	number = {4},
	journal = {Information visualization},
	author = {Soo Yi, Ji and Melton, Rachel and Stasko, John and Jacko, Julie A.},
	year = {2005},
	note = {ISBN: 1473-8716
Publisher: SAGE Publications Sage UK: London, England},
	pages = {239--256},
}

@article{leiva_enact_2019,
	title = {Enact: {Reducing} designer–developer breakdowns when prototyping custom interactions},
	volume = {26},
	shorttitle = {Enact},
	number = {3},
	journal = {ACM Transactions on Computer-Human Interaction (TOCHI)},
	author = {Leiva, Germán and Maudet, Nolwenn and Mackay, Wendy and Beaudouin-Lafon, Michel},
	year = {2019},
	note = {Publisher: ACM New York, NY, USA},
	pages = {1--48},
}

@inproceedings{martin_causette_2022,
	title = {Causette: user-controlled rearrangement of causal constructs in a code editor},
	shorttitle = {Causette},
	booktitle = {Proceedings of the 30th {IEEE}/{ACM} {International} {Conference} on {Program} {Comprehension}},
	author = {Martin, Alice and Magnaudet, Mathieu and Conversy, Stéphane},
	year = {2022},
	pages = {241--252},
}

@article{blouin_interacto_2021,
	title = {Interacto: {A} {Modern} {User} {Interaction} {Processing} {Model}},
	volume = {48},
	shorttitle = {Interacto},
	number = {9},
	journal = {IEEE Transactions on Software Engineering},
	author = {Blouin, Arnaud and Jézéquel, Jean-Marc},
	year = {2021},
	note = {Publisher: IEEE},
	pages = {3206--3226},
}

@inproceedings{hornbaek_what_2017,
	title = {What is interaction?},
	booktitle = {Proceedings of the 2017 {CHI} {Conference} on {Human} {Factors} in {Computing} {Systems}},
	author = {Hornbæk, Kasper and Oulasvirta, Antti},
	year = {2017},
	pages = {5040--5052},
}

@article{dimara_what_2019,
	title = {What is interaction for data visualization?},
	volume = {26},
	number = {1},
	journal = {IEEE transactions on visualization and computer graphics},
	author = {Dimara, Evanthia and Perin, Charles},
	year = {2019},
	note = {ISBN: 1077-2626
Publisher: IEEE},
	pages = {119--129},
}

@article{sedig_interaction_2013,
	title = {Interaction design for complex cognitive activities with visual representations: {A} pattern-based approach},
	volume = {5},
	number = {2},
	journal = {AIS Transactions on Human-Computer Interaction},
	author = {Sedig, Kamran and Parsons, Paul},
	year = {2013},
	note = {ISBN: 1944-3900},
	pages = {84--133},
}

@inproceedings{chi_operator_1998,
	title = {An operator interaction framework for visualization systems},
	isbn = {0-8186-9093-3},
	booktitle = {Proceedings {IEEE} {Symposium} on {Information} {Visualization} ({Cat}. {No}. {98TB100258})},
	publisher = {IEEE},
	author = {Chi, Ed Huai-hsin and Riedl, John T.},
	year = {1998},
	pages = {63--70},
}

@article{satyanarayan_reactive_2015,
	title = {Reactive vega: {A} streaming dataflow architecture for declarative interactive visualization},
	volume = {22},
	number = {1},
	journal = {IEEE transactions on visualization and computer graphics},
	author = {Satyanarayan, Arvind and Russell, Ryan and Hoffswell, Jane and Heer, Jeffrey},
	year = {2015},
	note = {ISBN: 1077-2626
Publisher: IEEE},
	pages = {659--668},
}

@article{schulz_design_2013,
	title = {A design space of visualization tasks},
	volume = {19},
	number = {12},
	journal = {IEEE Transactions on Visualization and Computer Graphics},
	author = {Schulz, Hans-Jörg and Nocke, Thomas and Heitzler, Magnus and Schumann, Heidrun},
	year = {2013},
	note = {ISBN: 1077-2626
Publisher: IEEE},
	pages = {2366--2375},
}

@article{rubab_examining_2021,
	title = {Examining interaction techniques in data visualization authoring tools from the perspective of goals and human cognition: a survey},
	volume = {24},
	journal = {Journal of Visualization},
	author = {Rubab, Sadia and Tang, Junxiu and Wu, Yingcai},
	year = {2021},
	note = {ISBN: 1343-8875
Publisher: Springer},
	pages = {397--418},
}

@article{zong_animated_2022,
	title = {Animated {Vega}-{Lite}: {Unifying} {Animation} with a {Grammar} of {Interactive} {Graphics}},
	volume = {29},
	number = {1},
	journal = {IEEE Transactions on Visualization and Computer Graphics},
	author = {Zong, Jonathan and Pollock, Josh and Wootton, Dylan and Satyanarayan, Arvind},
	year = {2022},
	note = {ISBN: 1077-2626
Publisher: IEEE},
	pages = {149--159},
}

@online{mouseEvent_Will,
	title = {How to create Mouse Events for D3},
	author = {William Liu},
	url = {http://bl.ocks.org/WilliamQLiu/76ae20060e19bf42d774},
	note = {Available at \url{http://bl.ocks.org/WilliamQLiu/76ae20060e19bf42d774}},
	year = 2004
}
